\documentclass[twocolumn,showpacs,graphicx,superscriptaddress,longbibligraphy]{revtex4}% Physical Review B
\usepackage{hyperref}
\usepackage{graphicx}% Include figure files
\usepackage{dcolumn}% Align table columns on decimal point
\usepackage{bm}% bold math
\usepackage[cmex10]{amsmath}
\usepackage[utf8]{inputenc}
\usepackage{float}
\usepackage{amssymb}
\usepackage[T1]{fontenc}
\usepackage{natbib}
 \usepackage{booktabs}
 \usepackage{multirow}
\usepackage{epsfig}

\begin{document}

%\title{ Co/Mo superlattices: the magnetoresistance, interlayer coupling  and magnetization dynamics}
\title{ Structural, magnetostatic, and magnetodynamic studies of Co/Mo-based uncompensated synthetic antiferromagnets 
%\\
%or
%\\
 %Structural, magnetostatic and magnetodynamic studies of Co/Mo-based synthetic ferrimagnets 
%\\
%or
%\\
%Structural, magnetostatic and magnetodynamic studies of Co/Mo superlattices in the presence of interlayer coupling
}

\author{Piotr Ogrodnik}
\email{piotr.ogrodnik@pw.edu.pl}
%\affiliation{University of Michigan, Department of Electrical Engineering and Computer Science, Ann Arbor, MI 48109, USA}
\affiliation{AGH University of Science and Technology, Department of Electronics, al. A. Mickiewicza 30, 30-059 Krakow, Poland}
\affiliation{Warsaw University of Technology, Faculty of Physics, ul. Koszykowa 75, 00-662 Warsaw, Poland}
%\altaffiliation[permanent address: ]{Warsaw University of Technology, Faculty of Physics, , ul. Koszykowa 75, 00-662 Warszawa, Poland}

%\author{Francesco Antonio Vetr\`{o}}
%\affiliation{\'{E}cole Polytechnique F\'{e}d\'{e}rale de Lausanne - EPFL, Institute of Physics, Station 3 , CH-1015 Lausanne, Switzerland}
\author{Jaros\l{}aw Kanak}
\affiliation{AGH University of Science and Technology, Department of Electronics, al. A. Mickiewicza 30, 30-059 Krakow, Poland}

\author{Maciej Czapkiewicz}
\affiliation{AGH University of Science and Technology, Department of Electronics, al. A. Mickiewicza 30, 30-059 Krakow, Poland}

\author{S\l{}awomir Zi\k{e}tek}
\affiliation{AGH University of Science and Technology, Department of Electronics, al. A. Mickiewicza 30, 30-059 Krakow, Poland}

\author{Aleksiej Pietruczik}
\affiliation{Institute of Physics, Polish Academy of Sciences, Aleja Lotnik\'{o}w 32/46, PL-02668 Warsaw, Poland}

\author{Krzysztof Morawiec}
\affiliation{Institute of Physics, Polish Academy of Sciences, Aleja Lotnik\'{o}w 32/46, PL-02668 Warsaw, Poland}

\author{Piotr D\l{}u\.zewski}
\affiliation{Institute of Physics, Polish Academy of Sciences, Aleja Lotnik\'{o}w 32/46, PL-02668 Warsaw, Poland}

\author{Krzysztof Dybko}
\affiliation{Institute of Physics, Polish Academy of Sciences, Aleja Lotnik\'{o}w 32/46, PL-02668 Warsaw, Poland}

\author{Andrzej Wawro}
\affiliation{Institute of Physics, Polish Academy of Sciences, Aleja Lotnik\'{o}w 32/46, PL-02668 Warsaw, Poland}

%\author{Jakub Chęci\'{n}ski}
%\affiliation{AGH University of Science and Technology, Department of Electronics, Al. Mickiewicza 30, 30-059 Krak\'{o}w, Poland}
%\affiliation{AGH University of Science and Technology, Faculty of Physics and Applied Computer Science, Al. Mickiewicza 30, 30-059 Krak\'{o}w, Poland}

\author{Tomasz Stobiecki}
\affiliation{AGH University of Science and Technology, Department of Electronics, al. A. Mickiewicza 30, 30-059 Krakow, Poland}
\affiliation{AGH University of Science and Technology, Faculty of Physics and Applied Computer Science, Aleja Mickiewicza 30, 30-059 Krakow, Poland}

%\author{J\'{o}zef Barna\'{s}}
%\affiliation{Adam Mickiewicz University, Faculty of Physics, ul. Umultowska 85, 61-614 Poznań, Poland}
%\affiliation{Institute of Molecular Physics, Polish Academy of Sciences, ul. M. Smoluchowskiego 17, 60-179 Poznań, Poland}

%\author{Jean-Philippe Ansermet}
%\affiliation{\'{E}cole Polytechnique F\'{e}d\'{e}rale de Lausanne - EPFL, Institute of Physics, Station 3 , CH-1015 Lausanne, Switzerland}

\date{\today}% It is always \today

\begin{abstract}
In this work, we comprehensively investigate and discuss the structural, magnetostatic, dynamic, and magnetoresistive properties of epitaxial Co/Mo superlattices. The magnetization of the Co sublayers is coupled antiferromagnetically with a strength that depends on the thickness of the nonmagnetic Mo spacer. The magnetization and magnetoresistance hysteresis loops clearly reflect interlayer exchange coupling and the occurrence of  uniaxial magnetic anisotropy induced by the strained Co sublayers.
%f the two-fold in-plane magnetic anisotropy. The anisotropy is modified when Co layers are sufficiently thick or thin to form hcp or fcc structures, respectively.
%tuned by the thickness of the component Co layers w . 
Upon accounting for a deviation of the sublayer thicknesses from the nominal value, theoretical modeling, including both micromagnetic and macrospin approaches, precisely reproduces experimental magnetic hysteresis loops, magnetoresistance curves, and ferromagnetic resonance dispersion relations. The Mo spacer thickness as a function of the interlayer magnetic coupling is determined as a fitting parameter by  modeling  the experimental results.
\end{abstract}


%Use showkeys class option if keyword
\maketitle

%\linenumbers
\section{Introduction}
Magnetic superlattices arising from the abundant contribution of atoms forming interfaces and from the  nonmagnetic layers being thinner  than the charge- and spin-transport characteristic lengths offer enhanced or completely new properties in comparison with the bulk materials. In particular, such multilayers have attracted a great deal of attention after the discovery of  interlayer exchange coupling  \cite{grunberg1986layered,majkrzak1986cf,salamon1986long}, which was observed in systems magnetized both in the layer plane  \cite{carbone1987antiparallel} or perpendicularly to the layer plane  \cite{grolier1993unambiguous}. The interlayer coupling was explained within the framework of Ruderman--Kittel--Kasuya--Yosida (RKKY) theory \cite{bruno1991oscillatory}, with quantum interference resulting from spin-dependent reflections of Bloch waves at the paramagnet-ferromagnet interfaces being due to confinement by ultrathin layers  \cite{bruno1995theory} or dipolar interactions. Due to this type of coupling, the alignment of magnetization was usually constrained to collinear or perpendicular configurations in the magnetic sublayers. The discovery of  interlayer exchanged-coupled systems was followed by that of a markedly enhanced spin-dependent transport effect called ``giant magnetoresistance'' (GMR) \cite{baibich1988giant,camley1989theory}. 

A strong antiferromagnetic coupling between two ferromagnetic layers separated by a thin nonmagnetic spacer (e.g., Ru) enabled the fabrication of synthetic antiferromagnets (SAFs) \cite{gomez2018synthetic,liu2018synthetic,chatterjee2018novel}. 
Recently,   multilayer SAF systems combined with a metal layer exhibiting strong spin-orbit interactions such as Pt/\(\text{Fe}_{20}\text{Ni}_{80}\)/Ru/\(\text{Fe}_{20}\text{Ni}_{80}\) \cite{chaudhuri2018electrical} or Pt/(Co/Pd)/Ru/(Co/Pd) \cite{zhangSOT} were used to demonstrate, particularly in patterned devices, an electrically detected spin-orbit torque ferromagnetic resonance (SOT-FMR) and magnetization switching between high- and low-resistance configurations. In addition, Wu \textit{et al.} \cite{wuSOT}  recently  demonstrated  SOT-induced field-free magnetization switching in an  exchange-coupled i-CoFeB/Mo(wedged)/p-CoFeB structure, where the letter ``i'' (``p'') stands for in-plane (perpendicular) magnetic anisotropy.

The magnetostatic properties of superlattices, with regular or more sophisticated structures, can be also analyzed by dynamic techniques, such as by ferromagnetic resonance (FMR) \cite{farle1998ferromagnetic,lindner2003ferromagnetic,khodadadi2017interlayer,rementer2017tuning,bezerra1999self,sousa2019ferromagnetic}.
Such  research seems to be particularly important provided the increase and tunability of FMR frequencies permit the design of novel ultrafast magneto-electronic devices \cite{li2018stress}.
Of the many types of superlattices, the heavy-metal/ferromagnetic (HM/FM)  systems are  investigated in particular  because they exhibit interfacial Dzyaloshinsky--Moriya interactions (DMIs) and DMI-related phenomena (e.g., the  formation  of magnetic skyrmions) \cite{hrabec2014,fert2017magnetic}. 
Although the Co/Mo system belongs to the HM/FM group, it is still rather poorly recognized. 

Current studies  mainly focus on how structural evolution correlates with magnetic properties \cite{zhao2003irradiation,yang2004evolution,yang2002structural,wang1991co,houserova2005phase,guo2003ab}. In addition, coupling effects have been reported in  several papers. For example, Parkin showed that oscillatory exchange coupling was a typical feature for numerous transition-metal spacers \cite{parkinCoMo}. The coupling strength increases both with the electron occupation of the \(d\) shell and along each column in the periodic table. In addition,  well-defined coupling with induced in-plane anisotropy was observed in  sputter-deposited multilayers \cite{shalyguina2006magneto}, and the structural and electronic factors affecting the coupling were analyzed theoretically for  transition-metal spacers, including Mo \cite{koelling1994magnetic}. However, this research did not quantitatively address the interlayer coupling strength.

In  previous work we  thoroughly investigated the magnetic properties (magnetic ordering and reversal processes) of Co/Mo/Co trilayers, both deposited on and covered by Mo or Au  layers. The Au buffer invokes isotropic behavior of the Co/Mo/Co trilayers. For  a thinner Co layer, the system displays perpendicular magnetization with oscillations between ferromagnetic (FM) and antiferromagnetic (AFM) alignments with \(d_{\text{Mo}}\), with the coupling field reaching as high as 3 kOe \cite{kurant2019}. For thicker \(d_{\text{Co}}\),  canted magnetization tuned by the coupling strength is observed. Conversely, due to a different crystalline symmetry and a lattice mismatch at the interfaces, the Mo buffer induces anisotropic strains that result in  in-plane twofold magnetic anisotropy \cite{wawro2017engineering}. Moreover, the in-plane magnetized trilayers exhibit an AFM interlayer coupling in the Mo spacer thickness (\(d_{\text{Mo}}\)) ranging from 0.5 to 1.0 nm. This coupling can be tuned or switched to FM type by ion irradiation \cite{wawro2017modifications,wawro2018magnetic}. Such an intended varying of the interlayer coupling enables periodic modification of the magnetization on the nanometer scale, which, in turn, allows  Co/Mo multilayer systems to be fabricated as switchable magnonic crystals \cite{wawro2018magnetic}. More-detailed synchrotron research on Co/Mo multilayered systems revealed a possible contribution of induced magnetic moments at Mo atoms of the spacer to the observed coupling, as reported recently  \cite{wawro2018momag}. 
Furthermore, it has been shown that Fe dopants in the Mo spacer suppress the interlayer coupling strength \cite{ukrsvekloi2019}.  

In the present work we investigate the structural, magnetic static and dynamic, as well as field-dependent transport properties of Co/Mo epitaxially grown superlattices with a greater number of bilayers. The magnetization of the Co sublayers  couples antiferromagnetically, forming the SAF structure.
To date, Parkin \cite{parkinCoMo} has produced the only publication available that discusses  interlayer exchange coupling in Co/Mo multilayers and considers the superlattice with 16 bilayers. In his work,  Co/Mo bilayers were sputter deposited, and  no detailed structural or dynamical studies are available for such systems. We show herein that the symmetry and lattice-parameter mismatches at the interfaces generate anisotropic strain in the Co layers. As a consequence,  magnetic anisotropy is induced in the sample plane with two mutually orthogonal  axes of magnetization (easy and hard). Thus, the superlattices are characterized by both tunable interlayer coupling and in-plane anisotropy, which strongly defines the magnetization orientation. In addition, we consider the layer thickness and distribution of magnetic parameters, which is an immanent feature of real multilayers and is usually not accounted for in interpretations of acquired results. This experimental study is supported by micromagnetic and macrospin numerical simulations of magnetostatic and magnetodynamic SAF properties. 

The paper is organized as follows:  Section \ref{sec:prepar} provides details of sample fabrication, 
and Sec. \ref{sec:experyment} explains the experimental techniques used to characterize the samples.  Section \ref{sec:struktura} analyzes the structural data from  high-angle x-ray diffraction (XRD) and low-angle x-ray reflectometry (XRR) measurements.  Section \ref{sec:magstudy} describes the magnetic properties of Co/Mo superlattices and is divided into several subsections: It starts with a presentation of the theoretical macrospin model in Sec. \ref{sec:makro}. The macrospin model has been validated by OOMMF micromagnetic calculations \cite{donahue1999oommf}, which are briefly described in Sec. \ref{sec:mikro}.
The results obtained from vibrating sample magnetometry (VSM) and four-probe magnetoresistance (MR) measurements are presented in Sec. \ref{sec:vsm}.
Section \ref{sec:dynamic} discusses in detail the magnetic dynamics in terms of FMR resonant modes, while the related results on interlayer coupling are given in Sec. \ref{sec:iec}. Finally, Sec.\ref{sec:sumary} concludes and summarizes the paper.

\section{Sample preparation\label{sec:prepar}}
Epitaxial Co/Mo superlattices were deposited onto (11-20)-oriented sapphire wafer substrates  by using a molecular-beam epitaxy  system (Prevac). All samples contained five Co/Mo bilayers in the following structures:
\(\text{\textit{S}/V(2.5)/Mo(0.6)/[Co(2.1)/Mo(0.6)}]\times \text{4/Co(2.1)/Mo(0.6)/V(3.0)}\) (called the  ``C1 coupled sample''), 
\(\text{\textit{S}/V(2.5)/Mo(0.6)/[Co(3.5)/Mo(0.7)}]\times \text{4/Co(3.5)/Mo(0.6)/V(3.0)}\) (called the ``C2 sample''),
and \(\text{\textit{S}/V(2.5)/Mo(0.6)/[Co(1.5)/Mo(0.9)}]\times \text{4/Co(1.5)/Mo(0.6)/V(3.0)}\) [called the ``weakly coupled (WC) sample''].
The  nominal thicknesses  given above are in nanometers, and \(S\) denotes a sapphire substrate (\(\text{Al}_2 \text{O}_3\)).   
The seed layer was thin V (110) because its crystalline structure is compatible with the substrate.
%The applied thin $V$(110) buffer's crystalline structure is compatible to the substrate's one.
%Therefore, the thin $V$ buffer has been be deposited instead of much thicker $Mo$ layer that would be required for a high quality growth of $Co$/$Mo$ multilayer structure. In addition, the $V$ layer thickness was optimal in terms of minimizing the short-circuit effects in magneto-transport measurements in the current-in-plane (CIP) configuration. 
The outer Co layers were also surrounded by  thin Mo  layers on the bottom and top to maintain the same type of interfaces. The entire stack was then covered again by a V capping layer. Except for the V buffer (deposited at room temperature and then annealed at 500\,\(^\circ\)C for 3 h), the remaining parts of the multilayered structures were deposited at room temperature. The component materials were evaporated from  electron guns at rates  of 0.05 nm/s, as measured by a quartz balance and a Hiden cross beam source. The crystalline structure was monitored \textit{in situ} by reflection high-energy electron diffraction. 

\section{Experiment\label{sec:experyment}}
The sample structure was characterized by using a high-resolution X’Pert–MPD diffractometer with a Cu anode. The samples were analyzed by using XRR and XRD. The diffractometer was equipped with an Euler cradle stage that allowed the sample to be rotated around the axis perpendicular to its surface and to be tilted from horizontal to vertical. This configuration allowed us to measure \(\phi\) scans (sample rotation) at fixed \(2\theta\) and \(\psi\) (sample tilt)  to determine the orientation of the crystallites with respect to the sample plane. Additionally, the Euler cradle allowed us to make  grazing-incidence x-ray diffraction  measurements, which provided  information on interplanar distances in the direction parallel to the sample surface.  

The high-angle annular dark-field (HAADF) scanning transmission electron microscopy study was done by using a Titan Cubed 300  microscope operating at 300 kV and equipped with an energy-dispersive x-ray (EDX) spectrometer. The cross-sectional lamellas were fabricated by using a focused Ga  ion beam  in a HeliosNanoLab system. Prior to preparing the lamellas,  a Pt layer was deposited by using an electron gun to protect the  sample surface from sputtering by Ga ions.

The magnetization hysteresis loops were measured in a high and low magnetic field by using  VSM  to  determine a saturation magnetization and characteristic loops typical of coupled magnetic layers.

The room-temperature MR  measurements were done on unpatterned, as-grown samples. Standard four-probe in-line contact alignment was implemented by using spring-type pins connected directly to the sample surface. This configuration is commonly called ``current in plane.''  The magnetic field applied in the sample plane was rotated with respect to the direction of the current.

The FMR spectra were acquired by using a HP 8720C vector network analyzer (VNA). Samples were placed face down on a 50-\(\Omega\)-matched coplanar waveguide  with a line width of 500 \(\mu\)m. Microwave transmission was measured at a constant frequency (ranging from 5 to 16 GHz) by sweeping the in-plane external magnetic field. The analysis of the relatively rich FMR spectra allowed us to determine the dispersion relations \(f(H_{\text{res}}), \) which are broadly discussed in Sec. \ref{sec:dynamic}.

\section{Structural studies\label{sec:struktura}}

This section first  shows that the considered Co/Mo SAF  forms a high-quality superlattice with well-distinguished sublayers with sufficiently sharp interfaces. Next, the thicknesses of all sublayers are determined with the help of XRR and XRD measurements and relevant simulations. Last but not least, the origin of the in-plane uniaxial magnetic anisotropy in Co layers is discussed and related to the distortion of hexagonal Co cells caused by the strain at the Co/Mo interfaces. 

\begin{figure}[h]
\includegraphics[width=8.6cm]{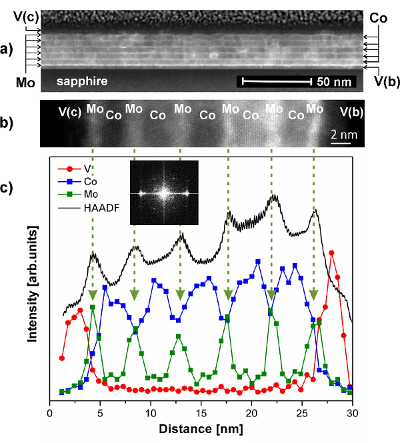}
\caption{TEM analysis of  sample C2: (a) Low-magnification TEM image of  Co/Mo multilayer. (b) High-resolution image with visible atomic layers with  V buffer [V(b)]  located on the right side. (c) In-depth EDX spectroscopy profile of chemistry of multilayer components: Co (blue), Mo (green), V (red),  and high-angle annular dark-field section (black). The inset shows a Fourier transform across the whole multilayer shown in panel (b).}
\label{fig:TEM}
\end{figure}

Figure \ref{fig:TEM}(a) shows a low-magnification TEM image of the C2 multilayer.  AFM interlayer magnetic coupling occurs in this sample and is discussed later in this work. The layered Co/Mo structure is composed of five Co layers, separated distinctly by the Mo spacers. The Mo layers are visible in the brightest color. The whole layered stack is deposited on the flat surface of the V buffer [denoted  V(b)] and also capped by a V film [V(c)]. The grainy structure in the upper part of the image illustrates the Pt covering layer that was deposited  to prepare a sample slice for  observation by transmission electron microscopy (TEM). The Mo layers seem to be continuous and relatively smooth. The columnar growth of Co and Mo layers results  in greater roughness for the  higher Mo layers. The presence of continuous Mo spacers is confirmed by a negligible influence of pinholes on AFM coupling, as discussed later in this work.

A closer look into the internal structure of the Co/Mo multilayer reveals a parallel alignment of atomic layers expanding across the layered structure of the sample [Fig. \ref{fig:TEM}(b)]. According to data from reflection high-energy electron diffraction  (data not shown) and XRD  (discussed later in this work), this observation confirms the epitaxial growth of the sample. The atomic planes parallel to the sample surface  consist of  bcc V (110), bcc Mo (110), and closed-packed hcp Co. %(due to different from the bulk-like growth is not possible to interpret unequivocally the $Co$ lattice as hcp or fcc).
The parallel alignment of the atomic layers across the entire sample is confirmed by the distinct spots in the Fourier transform shown in the inset of Fig. \ref{fig:TEM}(c), which presents the details of the multilayered sample structure. 
The in-depth chemical profile obtained from EDX spectroscopy is correlated with the TEM image. The maxima in the V profile correspond to the cap layer (left) and the buffer (right), and the  intensity oscillations in the Co and Mo signal  are very distinct. The lower-intensity signals are from the very thin (four atomic layers) Mo spacer. Electrons from the e-beam centered at the Mo spacer are scattered in the sample and excite  Co atoms in the vicinity of the interfaces. Details of the in-depth profiles are provided  by high-angle annular dark-field measurements (black line). The fine structure reflects atomic layers of the component films. Finally, the broad peaks correspond to the Mo spacers.

\begin{figure}[h]
\includegraphics[width=8.6cm]{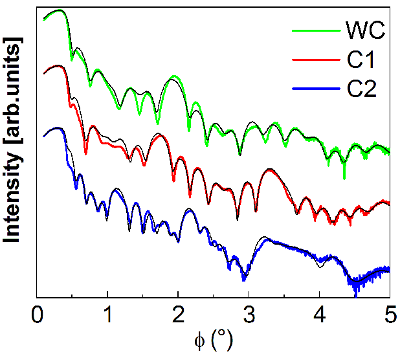}
\caption{XRR profiles of WC (green line), C1 (red line), and C2 (blue line) samples and corresponding fits (black lines).}
\label{fig:xrrprofile}
\end{figure}

The layered structure of the samples were investigated by using the XRR technique. Figure \ref{fig:xrrprofile} shows reflectivity curves and associated  numerical fits for the C1, C2, and WC samples. 
The fitting parameters show that the rms roughness at the Co/Mo interfaces increases  with the deposition of consecutive layers from 0.2  to 0.5 nm. The thicknesses of the individual Co and Mo layers are collected in Table \ref{tab:grubosci}. 
The thickness determined from the simulated XRR profiles  deviates from the assumed nominal thickness of the component layers. The actual measured quantities are used in numerical simulations of magnetostatic and magnetodynamic properties discussed later in this work.
\begin{table}[h!]
\centering
%\small
\caption{Thicknesses of  component layers (in nm unless stated otherwise) obtained from XRR measurements.
 \label{tab:grubosci}}
%\begin{tabular}{ccccccc}
\begin{tabular}{ccccccccccccc}
\hline\hline
& \text{Layer} & &\text{C1} & & \text{C2} & & \text{WC}\\

\hline

\text{Substrate:}& \(\rm{Al_2 O_3}\) & & 0.6 mm & & 0.6 mm & & 0.6 mm \\
  &V & & 2.78 & & 2.46 & & 2.53 \\
  &Mo& & 0.67 & & 0.61 & & 0.81 \\ 
  &Co & & 2.14 & & 3.34 & & 1.48 \\
& Mo & & 0.52 & & 0.77 & & 0.89 \\
  &Co & & 2.01 & & 3.60 & & 1.52 \\
  &Mo & & 0.60 & & 0.56 & & 0.99 \\
  &Co & & 2.03 & & 3.70 & & 1.03 \\
  &Mo & & 0.55 & & 0.57 & & 0.89 \\
  &Co & & 2.13 & & 3.58 & & 1.77 \\
  &Mo & & 0.68 & & 0.76 & & 1.05 \\
  &Co & & 2.21 & & 3.20 & & 1.75 \\
  &Mo & & 0.64 & & 0.68 & & 0.52 \\
\text{Capping:} &V & & 3.33 & & 3.00 & & 3.33 \\
    \hline\hline
\end{tabular}
\end{table}

\begin{figure}[h]
\includegraphics[width=7.5cm]{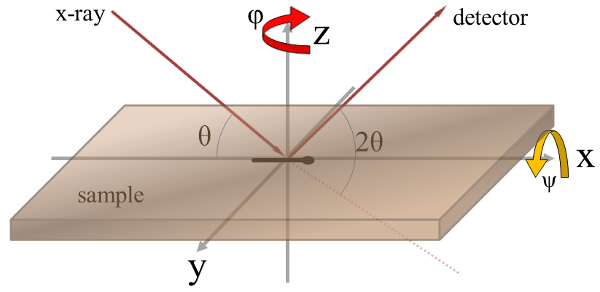}
\caption{Scheme of the XRD scans: \(\phi\) (rotation around the \(z\) axis) and \(\psi\) (rotation around the \(x\) axis). The black line with a dot  on the sample indicates the sample's initial position (\(\phi = 0\)).}
\label{fig:schematxrd}
\end{figure}

\begin{figure}[h]
\includegraphics[width=7.5cm]{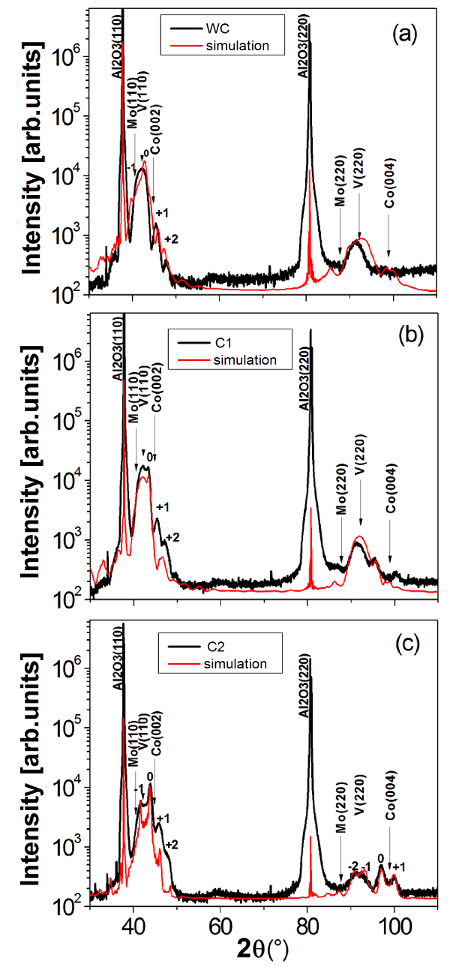}
\caption{XRD \(\theta\text{-}2\theta\) spectra (black lines) for multilayered samples WC (a), C1 (b), and C2 (c). Red lines are the simulation of the diffraction profiles.}
\label{fig:figura2}
\end{figure}

The crystalline structure of the component layers was also examined with XRD. Figure \ref{fig:schematxrd} shows a schematic configuration of the XRD %$2\theta$ 
measurements at different \(\phi\) and \(\psi\) angles. A change in the angle \(\phi\)  corresponds to a rotation of the sample around the \(z\) axis, which is perpendicular to the sample plane, whereas the \(\psi\) angle is a tilt of the sample around the \(x\) axis. The black line with a dot  on the sample defines the initial position of the sample for \(\phi = 0^\circ\). Figure \ref{fig:figura2} shows \(\theta\text{-}2\theta\) (at \(\phi=0^\circ\) and \(\psi=0^\circ\)) XRD spectra from C1, C2, and WC samples.
 Additional peaks appear on the right side of the two very strong peaks that originate from the sapphire (\(\rm Al_2 O_3\)) substrate. The arrows indicate the positions of peaks for bulk-like structural bcc V (1 1 0), hcp Co (0 0 2), and bcc Mo (1 1 0)  and their second-order peaks. However, their positions do not coincide with those of superlattice peaks in the measured diffraction profile. To explain the origin of these peaks, the \(\theta\text{-}2\theta\) diffraction model was used to simulate the periodic structure of the artificial superlattice  \cite{kanak2013}. The sample structures were simulated by using the thicknesses of the individual layers obtained from the XRR measurements.  The calculated \(\theta\text{-}2\theta\) diffraction profiles are plotted in Fig. \ref{fig:figura2} as red lines. The simulations  show clearly that the  peaks in the diffraction profiles are not structural peaks from Mo and Co layers but originate from the Co/Mo superlattices and  reproduce in entirety the experimental patterns. These results confirm the good planar growth of the layered structure throughout the stack.

\begin{figure}[h]
\includegraphics[width=7.5cm]{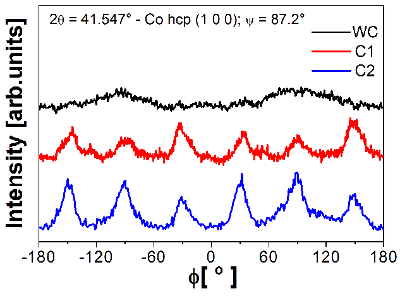}
\caption{XRD \(\phi\) scans for multilayer samples  C1, C2, and WC measured at  \(2\theta = 41.6^{\circ}\) and \(\Psi = 2\theta = 87.2^{\circ}\).}
\label{fig:figura3}
\end{figure}
Figure \ref{fig:figura3} shows \(\phi\) scans for the C1, C2, and WC samples at  \(2\theta = 41.6^{\circ}\), which corresponds to  hcp Co (1 0 0) planes. The angle \(\psi\) was set to \(87.2^\circ\) (i.e., near \(90^\circ\)), which corresponds to the angle between the hcp Co (1 0 0) and (0 0 2) planes. The angle \(87.2^\circ\) was chosen because, at  higher angles, the peak intensities were strongly suppressed because the x-ray beam was partially blocked by the sample edge. The C1 and C2 samples exhibit six pronounced peaks  from Co hcp (1 1 0) planes, whereas these peaks do not appear for the WC sample. The six peaks in the diffraction profiles originate from the sixfold symmetry of the Co hexagonal structure. Unlike the C1 and C2 structures, the WC multilayer does not have sixfold symmetry. The  Co sublayers in the WC sample are the thinnest layers in comparison with the other samples. Due to lattice mismatch at the Co/Mo interfaces, the  Co layer in the initial growth stage has poor crystalline structure (most probably  a mixture of hcp and fcc phases, with numerous stacking faults). In addition, the crystallinity of the Mo spacer, which affects the growth of the subsequent Co layer, depends on the crystalline structure of the  Co film underneath.  

\begin{figure}[h]
\includegraphics[width=7.5cm]{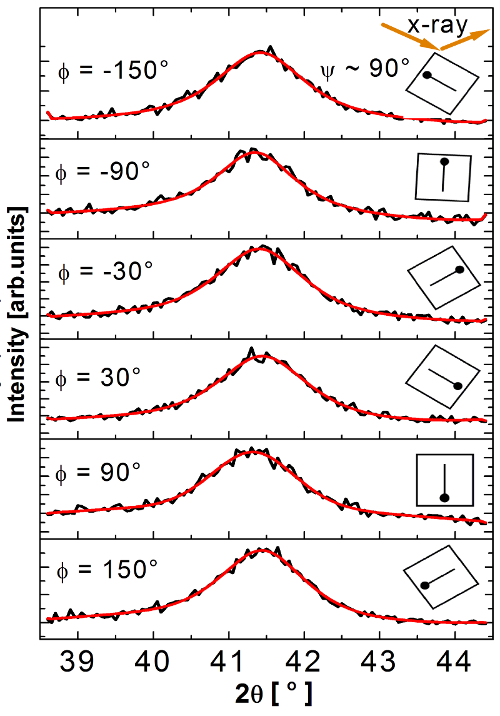}
\caption{XRD \(\theta\text{-}2\theta\) profiles (black lines) for C1 sample. Red lines are  fits to the diffraction profiles.}
\label{fig:figura4}
\end{figure}

\begin{figure}[h]
\includegraphics[width=7.5cm]{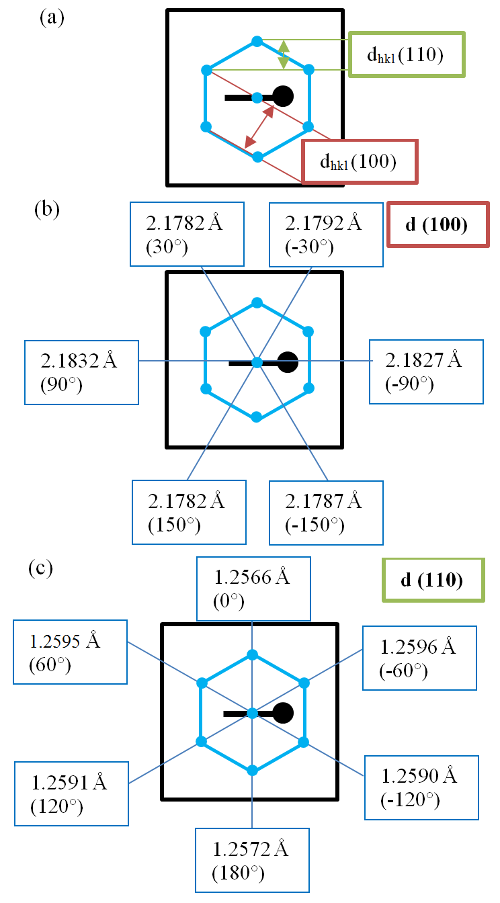}
\caption{Strain-induced deformation of the hexagonal Co cell in the C1 sample: (a) Definition of the interplanar distances, (b) azimuthal (\(\phi\)) dependence of \(d\)(100), and (c) \(d\)(110)  interplanar distances.}
%Schematic Co hexagonal cell in sample (a), interplanar distances d (100) (b) and d (110) (c) of Co hexagonal cell for C1 sample depending on the $\phi$ angle.}
\label{fig:hex}
\end{figure}

Figure \ref{fig:figura4} shows \(\theta\text{-}2\theta\) scans of the C1 sample taken around \(2\theta \approx 41.6^\circ\), which  corresponds to the Co hcp (1 0 0) planes. Measurements were taken at six maxima obtained from the  \(\phi\) scan (see Fig. \ref{fig:figura3}) with the sample tilted  by \(\psi \approx 90^\circ\) about the \(x\) axis. 
The insets in Fig. \ref{fig:figura4} show the \(\phi\) angles of the sample during the measurement. The angular positions of the peaks were determined from the fits to the peak profiles (red lines). % and collected in Table \ref{tab:tabelaxrd}. 
A similar analysis was made  for the Co (1 1 0) peak. Additionally, from the peak positions for different \(\phi\) angles, the interplanar distances of Co (1 0 0) and Co (1 1 0) were calculated, with  the detailed results  presented in Table \ref{tab:tabelaxrd}.
Figure \ref{fig:hex}(a) illustrates a Co hexagonal cell with the indicated interplanar spacings. Figures \ref{fig:hex}(b) and \ref{fig:hex}(c) show the interplanar spacings \(d\)(100) and \(d\)(110) in the respective directions. The interplanar spacing \(d\)(100) is larger along the direction marked by the  thick black pointer inside the cell than in the direction perpendicular to the pointer. Similarly, the distances \(d\)(110) are considerably larger in the angular directions  \(\phi = \pm 60^\circ,\pm 120^\circ\) than at \(\phi=0^\circ,180^\circ\). These results  show clearly that the hexagonal Co cells are stretched in the direction indicated by the black pointer. The same type of  Co-lattice deformation was reported by Prokop \textit{et al.} \cite{prokop2004}. 

\begin{figure}[h]
\includegraphics[width=7.5cm]{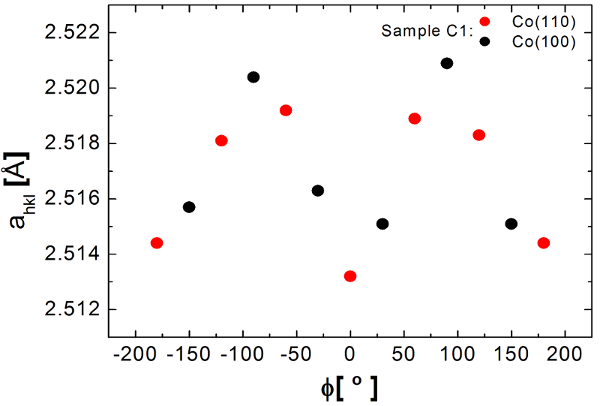}
\caption{Interplanar distances \(a_{hkl}\)  as a function of \(\phi\)    in Co hcp cell in C1 sample.}
\label{fig:distance}
\end{figure}

\begin{table}
\caption{Angular (\(\phi\)) dependence of \(2\theta\) and the calculated lattice parameters for C1 sample.
 \label{tab:tabelaxrd}}
\begin{tabular}{ccccccccc}
\hline
\hline
\multicolumn{4}{c}{Co (1 1 0)}  && \multicolumn{4}{c}{Co ( 1 0 0)} \\
\cline{1-4}\cline{6-9}
& & & & & & & \\
\textbf{\(\phi\) } & \(2\theta \) & \textbf{\(d ( 1 1 0)\) }& \textbf{\(a_{hkl}\)} && \textbf{\(\phi\) } & \(2\theta \) &  \textbf{\(d ( 1 0 0 )\) }& \textbf{\(a_{hkl}\) }\\
\((^\circ\)) & \((^\circ\)) & (nm) & (nm) && \((^\circ\)) & \((^\circ\)) & (nm) & (nm) \\

\hline
\(-150\) &  41.41 & 2.1787 & 2.5157& & \(-120\) & 75.44 & 1.2590 & 2.5181\\
\(-90\) &  41.33 & 2.1827 & 2.5204 && \(-60\) & 75.40 & 1.2596 & 2.5192\\
\(-30\) & 41.40 & 2.1792 & 2.5163 && 0 & 75.61 & 1.2566 & 2.5132\\
30 &  41.42 & 2.1782 & 2.5151 && 60 & 75.41 & 1.2595 & 2.5189\\
90 & 41.32 & 2.1832 & 2.5209 && 120 & 75.43 & 1.2591 & 2.5183\\
150 & 41.42 & 2.1782 & 2.5151 && 180 & 75.57 & 1.2572 & 2.5144 \\
\hline\hline
\end{tabular}
%\end{ruledtabular}
\end{table}
Figure \ref{fig:distance} shows the lattice constant versus \(\phi\)  calculated from the interplanar spacings \(d\)(100) and \(d\)(110).
The larger lattice constant is for \(\phi= -90^\circ\) and 90\(^\circ\). %that correspond to the sample easy axis of a magnetization. 
Similar measurements (data not shown) were carried out for the C2 sample and revealed the same relations.
How the determined in-plane distortion of the Co cell in the C1 and C2 samples affects the magnetic properties is discussed later in this work. The results suggest that uniaxial strain is present in the sample plane. Magnetoelastic coupling should give rise to  magnetic in-plane anisotropy. As discussed in the next sections, such anisotropy is clearly developed and imposes an easy-axis orientation of 90\(^\circ\). The same relation between the lattice deformation and magnetic anisotropy was reported in Ref. [\onlinecite{prokop2004}].

\section{Magnetic studies\label{sec:magstudy} }
\subsection{Theoretical model\label{sec:theory}}
This subsection  presents the macrospin and micromagnetic models that allow us to calculate SAF magnetization hysteresis loops and to analyze the magnetic dynamics of the considered structure in terms of resonance frequencies and resonance-mode intensities. 
\subsubsection{Macrospin\label{sec:makro}}
To effectively simulate  the system, the macrospin approach was applied. The resonance frequencies of the magnetic structures can be easily calculated by using the well-established Smit--Beljers theory \cite{smit1955}. However, here the proposed approach is more relevant to deal with superlattices consisting of several strongly coupled magnetic layers. The superlattice investigated in this paper consists of five coupled magnetic Co layers separated by nonmagnetic Mo layers of different thickness. We observe experimentally the magnetic response  from the dynamic behavior of the magnetic moments of each layer, which are described by five pairs of spherical angles (polar \(\theta_i\) and azimuthal \(\phi_i\)): 
\begin{equation}
\bm{M_i} = M_{S,i} [ \sin\theta_i \cos\phi_i,\, \sin \theta_i \sin\phi_i ,\, \cos\theta_i],
\end{equation}
where \(i\) corresponds to the \(i\)th cobalt layer. 
Therefore, the magnetization dynamics of the system is described by five coupled Landau--Lifshitz--Gilbert (LLG) equations,
\begin{equation}
\label{eq:LLGset}
\frac{d\bm{M_i}}{dt} = -\gamma_e \bm{M_i} \times \bm{H}_{\rm eff,i} + \frac{\alpha_g}{M_{S,i}} \bm{M_i} \times \frac{d\bm{M_i}}{dt},\\
\end{equation}
where \(\gamma_e\approx1.760859644\times 10^{11}\ \frac{\mathrm{rad}}{\mathrm{sT}}\) is the gyromagnetic ratio, and \(\alpha_g\)  is the Gilbert damping parameter for each layer. 
The value of the damping parameter (\(\alpha_g = 0.02\)) was determined from the widths of VNA-FMR resonant peaks from the WC sample (an example of such spectra is discussed in Sec. \ref{sec:dynamic}). In general, the damping in the Co layer may  vary considerably, even by one order of  magnitude \cite{barati}, depending on the  thickness of the Co and adjacent Mo layers. However, a smaller variation in \(\alpha_g\) has been reported for thicker layers. In our case, both Co and Mo layers are sufficiently thick to fix the value of  \(\alpha_g\)  for all samples considered herein without harming the reliability of the simulations.
The effective fields can be expressed as follows:  
\begin{equation}
\bm{H}_{\rm eff,i} = -\bm{\nabla}_{\theta_i,\phi_i} U , 
\end{equation}
where \(\bm{\nabla}_{\theta_i,\phi_i}\) are  the relevant gradients in spherical coordinates.
The total magnetic energy density \(U\) can be written as
\begin{eqnarray}
U(\theta_i,\phi_i) &=& \sum_{ij}  -K_i d_{Co_i}(\cos\phi_i \sin\theta_i)^2  -d_{Co_i} \bm{M_i} \cdot \bm{H}_{\rm ext} \nonumber \\
&&-d_{Co_i} \bm{M_i} \cdot \bm{H}_{dem,i} -  J_{ij} \bm{M_i} \cdot \bm{M_j},
\label{eq:energia}
 \end{eqnarray}
where \(K_i\) is the anisotropy constant of the \(i\)th Co layer,  \(\bm{H}_{\rm ext}\) is an external magnetic field,  \(\bm{H}_{{\rm dem},i}\) is the demagnetizing field within the layers, and \(J_{ij}\) is  the bilinear interlayer exchange-coupling energy constant.

To calculate the resonance frequencies, we assume small-amplitude oscillations, which allows us to linearize  Eq. (\ref{eq:LLGset}) around the energy minima at given external magnetic-field magnitudes. Provided small-amplitude oscillations are taken into account, the angular solutions of Eq. (\ref{eq:LLGset}) can be expressed as harmonic oscillations of magnetization angles in polar coordinates: 
\begin{eqnarray}
\theta_i(t)&=&\theta_{0,i} + \delta\theta_i (t) = \theta_{0,i} + \delta\theta_i e^{i(\omega_i t)},\\
\phi_i(t)&=&\phi_{0_i} + \delta\phi_i (t) = \phi_{0,i} + \delta\phi_i e^{i(\omega_i t)},
\label{eq:rozw}
\end{eqnarray}
where \((\theta_{0,i},\, \phi_{0,i})\) describe the equilibrium orientation of the magnetization in all magnetic layers (in the absence of a driving radio frequency magnetic field).
In general, the amplitudes \((\delta\theta_i, \,\delta\phi_i)\) are complex and include a phase shift between the magnetization and the time-dependent driving force (e.g., an AC external magnetic field). Also, the frequencies \(\omega_i\) are complex. Their real parts \(\omega_{R,i}\) correspond to resonance angular frequencies, whereas the imaginary parts \(\frac{1}{\omega_{I,i}}\) give the half-life of decaying free oscillations due to presence of effective damping. As done in Ref.~ \cite{Ogrodnik_2018},  Eq. (\ref{eq:LLGset}) can be rewritten in  matrix form as
\begin{equation}
\dot{\bm{\alpha}} = \bm{v}(\theta_i,\phi_i)^T,
\label{eq:llgpolar}
\end{equation}
where \(\bm{\alpha} = \left( \alpha_1 ,\ldots, \alpha_{10} \right)^T \equiv \left( \theta_1,\phi_1, \ldots, \theta_5,\phi_5 \right)^T \) is a 10-element column vector consisting of time derivatives of spherical angles given by Eq. (\ref{eq:rozw}). Note that components of \(\bm{\alpha}\) are indexed by \(k\) varying from 1 to 10.  
Thus, the LLG equation in polar coordinates has the general form
\begin{equation}
\dot{\bm{\alpha}} = \bm{v}(\theta_i,\phi_i)^T,
\label{eq:llgpolar}
\end{equation}
where \(\bm{v}\) is the right-hand-side (RHS) vector of the LLG equation. After linearization of \(\bm{v}\) with respect to small deviations in \(\theta_i\) and \(\phi_i\) from their stationary values, one can write  Eq. (\ref{eq:llgpolar}) in the form
\begin{equation}
\dot{\bm{\alpha}} = \hat{X}  \bm{\Gamma}(t) ,
\label{eq:llglinear}
  \end{equation}
where  \(\hat{X}\) is a \(10 \times 10\) matrix consisting of the derivatives of the RHS of Eq. (\ref{eq:llgpolar}) with respect to the angles \(\theta_i,\,\phi_i\) (i.e., \(X_{kj} \equiv \frac{\partial v_{k}}{\partial \alpha_j}\)), while \(\bm{\Gamma}(t)= \bm(\delta\alpha_1(t), \ldots, \delta \alpha_{10}(t) \bm )^T\) is a vector containing time-dependent angle differentials [i.e., \(\delta \alpha_1(t) \equiv \delta \theta_1(t)\), \(\delta \alpha_2(t) \equiv \delta \phi_1(t)\), etc.], as defined in Eq. (\ref{eq:rozw}). Next,  Eq. (\ref{eq:llglinear}) can be rewritten as an eigenvalue problem of the matrix \(\hat{X}\):
 \begin{equation}
  \left| \hat{X} - \omega \hat{I} \right| = 0.
  \label{eq:czesto}
 \end{equation}
The eigenvalues \(\omega_i\) determine five distinguished resonance (natural) angular frequencies of the system, \(\omega_{R,k} = \text{Re}\, \omega_i \). 
On the other hand, the general solution of  Eq. (\ref{eq:LLGset}) can be expressed by using a linear combination of eigenvectors corresponding to all eigenvalues \(\omega\) in the following form:
\begin{equation}
\label{eq:eigensolution}
\bm{\alpha}(t) = \sum_{k=0}^{10} c_k \bm{\mathrm{v}_k} e^{i\omega_k t} ,
\end{equation}
where \(\bm{\mathrm{v}}_k\) is a 10-element eigenvector corresponding to the eigenvalue \(\omega_k\), and \(c_k\) are the scalar coefficients that can be combined as one column vector \(\bm{c}\). 
Equation (\ref{eq:eigensolution}) describes the free damped oscillations that may occur when the magnetizations are initially tilted away from their minimum energy points by, for example, an external magnetic field. 
For \(t=0\), Eq. (\ref{eq:eigensolution}) can be expressed by using the scalar product 
\begin{equation}
\bm{\alpha}(0) = \bm{c}\cdot \bm{\mathrm{v}} ,
\end{equation}
so that the columns of the vector \(\bm{\mathrm{v}}\) are the eigenvectors \(\bm{\mathrm{v}}_k\). Thus, the components of \(\bm{c}\) can be derived by using the initial conditions for magnetization angles and their time derivatives; namely, 
\begin{equation}
\bm{\alpha}(0)= \sum_{k=0}^{10}  c_k \bm{\mathrm{v}_k}   = \bm{\alpha_{0}}
\label{eq:cond1}
\end{equation}
 and 
 \begin{equation}
 \bm{\dot{\alpha}} (0) =  \sum_{k=0}^{10}i\omega_k c_k \bm{\mathrm{v}_k}  =  \bm{v} (\bm{\alpha_{0}}),
\label{eq:cond2}
 \end{equation}
 where \(\bm{\alpha}_0 \equiv (\alpha_1,...,\alpha_{10}) = (\theta_{0,1},\phi_{0,1}, ... , \theta_{0,5},\phi_{0,5})\). 
By combining Eqs. (\ref{eq:cond1}) and (\ref{eq:cond2}), the formula for the mode-coefficient vector \(\bm{c}\) takes the form
\begin{equation}
\bm{c}=  \bm{v}(\bm{\alpha_0}) \cdot \bm{\mathrm{\tilde{v}}}^{-1} ,
\end{equation}
where \(\mathrm{\tilde{v}}_k \equiv (1+i \omega_k) \mathrm{v}_k \). The initial values of angles \(\bm{\alpha_0}\) can be determined by minimizing  the total magnetic energy (\ref{eq:energia}). 
The intensity \(\mathcal{I}_k\) of the given resonance mode (\(\omega_{k}\)) is defined as
\begin{equation}
\label{eq:intense}
\mathcal{I}_k \equiv \max\bm(\xi(t)\bm)-\min\bm(\xi(t)\bm),
\end{equation}
where
\begin{eqnarray}
\xi(t)&=&   \sum_{l=1}^{5}\bm{ M_{0,l}}+\Delta \bm{M_l} 
\label{eq:defint}\\
&\approx &\sum_{l=1}^{5} M_{S,l} \left\{ \cos \left[ \alpha_{0,2l} + \text{Re} \left( c_{k,2l} \cdot \rm{v}_{k,2l} e^{i\omega_{k} t }\right) \right] \right.  
\nonumber\\
&&\left. +\sin\left[\alpha_{0,2l} + \text{Re} \left( c_{k,2l} \cdot \rm{v}_{k,2l} e^{i\omega_{k} t } \right) \right]\right\}.\nonumber
\end{eqnarray}
%\mathcal{I}_k \equiv \max ( \Re \left( c_k \cdot \rm{v}_k e^{i\omega_{k} t} \right) )
The definition only accounts for the changes in the in-plane magnetization  (described by \(\phi_i\)). However, it is clearly consistent with the fast Fourier transform  procedure that we apply to the results of the micromagnetic simulation. 
Equation (\ref{eq:defint}) ensures that \(\mathcal{I}_k\)  is suppressed when, for example, adjacent antiparallel magnetization vectors oscillate both with the same phase. On the contrary,  \(\mathcal{I}_k\) is enhanced when both antiparallel magnetizations oscillate with the opposite phases. 

\subsubsection{Micromagnetics \label{sec:mikro}}
Equation (\ref{eq:intense}) allows us to compare macrospin-resonance-mode intensities with the micromagnetic ones. Thus we have performed micromagnetic simulations with use of the OOMMF package \cite{donahue1999oommf}. First, we modeled the hysteresis loops, then we calculated the dynamic magnetization  response of the structure upon excitation by an external field. The latter  gave us  the resonance frequencies of the system and the intensities of the resonance modes. Both static and dynamic micromagnetic simulations employ cuboid-shaped simulation cells with dimensions \(5\text{ nm} \times 5\text{ nm} \times 1.5\text{ nm}\). The exchange constant for all Co layers was taken as \(A_{\text{ex}} = 3 \times 10^{-11} \) J/m. The dynamics simulation scheme is similar to that used in our previous works \cite{Ogrodnik_2018,frankowskicomputer}:  it starts from the state of fully relaxed magnetization of the Co/Mo superlattice. Next,  to excite a dynamic response, a short magnetic-field pulse (of the order of 10 Oe) is applied, and the magnetization response of the system is processed by using a fast Fourier transform  to disclose the natural frequencies at a given external magnetic field.  
As shown in  Secs. \ref{sec:static} and \ref{sec:dynamic}, the micromagnetic simulations fully confirm the reliability of the macrospin model in both cases: coupled and WC Co/Mo multilayers. Next, these two models are applied  to comprehensively characterize coupled and WC Co/Mo superlattices.

\subsection{Magnetostatic characterization\label{sec:static}}

This subsection presents the experimental results from VSM [\(M(H\))] and MR [\(MR(H)\)] measurements of the three  SAF structures considered (i.e., samples C1, C2, and WC). To reproduce the observed relations and to determine the magnetic parameters, the macrospin model is fit to the experimental data. In particular, we estimate the magnitude of the coupling energy between Co layers through a Mo spacer. Next, the reversal processes in SAF are presented and analyzed through macrospin modeling. The micromagnetic simulations show the reliability of this macrospin approach.

\subsubsection{Hysteresis loops and magnetoresistance\label{sec:vsm}}

To gain insight into the magnetization-reversal process in the  structures considered,  VSM and MR  measurements were carried out. 
The hysteresis loops and MR dependencies were measured in the magnetic field applied in the sample planes along mutually orthogonal easy  and hard  axes being a 
consequence of magnetoelastic strain of the hcp cobalt cell induced at the Co/Mo interfaces (this effect is discussed in Sec. \ref{sec:struktura} and in Ref. \cite{wawro2017engineering}). 
The MR in Co/Mo multilayers usually takes very small values and does not exceed 0.1\%\ regardless of the presence of  relatively strong interlayer coupling \cite{ukrsvekloi2019,bloemen1996interlayer}. 
However, its magnetic-field dependence precisely reflects  all the magnetic features revealed by the \(M(H)\) hysteresis loops.
The \(MR(H)\) data turn out to be useful to determine the magnetic parameters of the SAF structures. The dominating contribution to the total MR is related to the GMR effect, which is very sensitive to the relative angle between the magnetizations of two neighboring magnetic layers. Therefore,  simultaneously fitting the macrospin model to the \(M(H)\) and \(MR(H)\) dependencies should result in   optimal and reliable macrospin magnetic parameters. However, this approach can be applied only to the C1 and C2 samples; the MR is completely suppressed in the WC sample, which may be related to the thickness of the Co layers \cite{shukh1994dependence} being less than those of  other samples.
%, and/or their mixed structures (with non-hcp contribution), as already discussed above.
For this reason, for the WC sample, only VSM data are used for the model fitting.

To model the experimental data presented in this paper, we used a self-consisting fitting of the macrospin model to the VSM hysteresis loops and MR relations in both the easy and hard directions. 
The orientation of the magnetic moments was calculated by minimizing the total magnetic energy of the system given by Eq. (\ref{eq:energia}). In the fitting procedure, we used the Co layers thicknesses determined by XRR measurements. The GMR value was calculated by using the standard formula \( R_{ij}(\theta_{ij}) = R_0 + \frac{\Delta R}{2} ( \cos \theta_{ij} - 1) \), where \(\Delta R \equiv R_{AP,ij} - R_{P,ij}\) is the difference in resistance between layers \(i\) and \(j\) in the parallel (P) or antiparallel (AP) state, and \(\theta_{ij}\) is the relative angle between magnetic (Co) layers \(i\) and \(j\)  and  depends implicitly on the external magnetic field \(H\). In the current-in-plane configuration, the MR originates from each pair of magnetic layers, so the overall GMR of the whole structure is calculated by averaging \(R_{ij}\) over all pairs at a given magnetic field [i.e., \( R(H) = \frac{R_{ij}}{4}\)]. Despite this simplification, the theoretical curves are sufficiently consistent with the experimental curves. The  optimal set of macrospin magnetic parameters determined from the VSM and MR data is listed in Table \ref{tab:tabela}.

\begin{table}
\centering
\small
\caption{Set of  macrospin magnetic parameters for C1, C2, and WC samples. The parameters are used in the theoretical plots of hysteresis loops and MR dependencies shown in Figs. \ref{fig:gmrcmacro}--\ref{fig:mikroVSM}.  \label{tab:tabela}}
%\begin{ruledtabular}
\begin{tabular}{cccccc}
\hline\hline

\text{Sample } & {Layer \(i\)}  & \(d_i\) [nm]  &  \(K_i\) [\(\frac{\text{kJ}}{\text{m}^3}\)]   & \(M_{S,i}\) [T] &  \(J_{i,j}\) [\(\frac{\text{mJ}}{\text{m}^2}\)] \\
\hline

\textbf{C1}  & Co 1  &  2.14  & 1.0  & 1.35  &  \\
      &  Mo  & 0.52 &   &   & \(-0.10\) \\
  & Co 2  &  2.01  & 1.0  & 1.34  &  \\
        &  Mo  & 0.60 &  &   & \(-0.18\) \\
  & Co 3  &  2.03  & 1.0  & 1.34  & \\
        &  Mo  & 0.55 &  &   & \(-0.20\) \\
  & Co 4  &  2.10  & 1.0  & 1.35  &  \\
        &  Mo  & 0.70 &  &   & \(-0.16\) \\
  & Co 5  &  2.21  & 1.0  & 1.36  &  \\
  \hline  
  \textbf{C2}  & Co 1  &  3.34  & 2.5  & 1.40  &  \\
      &  Mo  & 0.77 &   &   & \(-0.15 \)\\
  & Co 2  &  3.60  & 2.5  & 1.40  &  \\
        &  Mo  & 0.56 &  &   & \(-0.21\) \\
  & Co 3  &  3.70  & 2.5  & 1.40  &  \\
        &  Mo  & 0.57 &  &   & \(-0.22\) \\
  & Co 4  &  3.58  & 2.5  & 1.40  &  \\
        &  Mo  & 0.76 &  &   & \(-0.16\) \\
  & Co 5  &  3.20  & 2.5  & 1.40  &  \\
  \hline
    \textbf{WC}  & Co 1  &  1.48  & 0  & 1.30  &  \\
      &  Mo  & 0.86 &   &   & \(-0.012\) \\
  & Co 2  &  1.52  & 0  & 1.30  &  \\
        &  Mo  & 0.99 &  &   & \(-0.012\) \\
  & Co 3  &  1.03  & 0  & 1.10  &  \\
        &  Mo  & 0.89 &  &   &\(-0.015\) \\
  & Co 4  &  1.77  & 0  & 1.37  &  \\
        &  Mo  & 1.05 &  &   & \(-0.012 \)\\
  & Co 5  &  1.75  & 0  & 1.37  & \\
  \hline\hline 
\end{tabular}
%\end{ruledtabular}
\end{table}

\begin{figure}
\includegraphics[width=8.6cm]{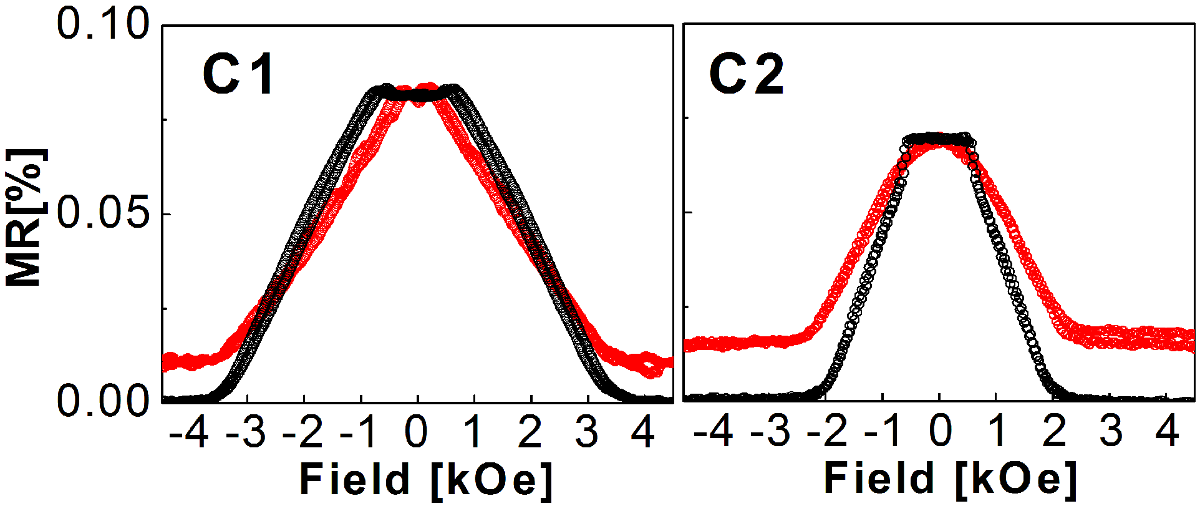}
\caption{Magnetoresistance of  samples with AFM-coupled magnetic layers: C1 (left panel) and C2 (right panel) measured with magnetic field applied along the hard axis (red points) and easy axis (black points).  
}
\label{fig:gmrc}
\end{figure}

\begin{figure}
\includegraphics[width=8.6cm]{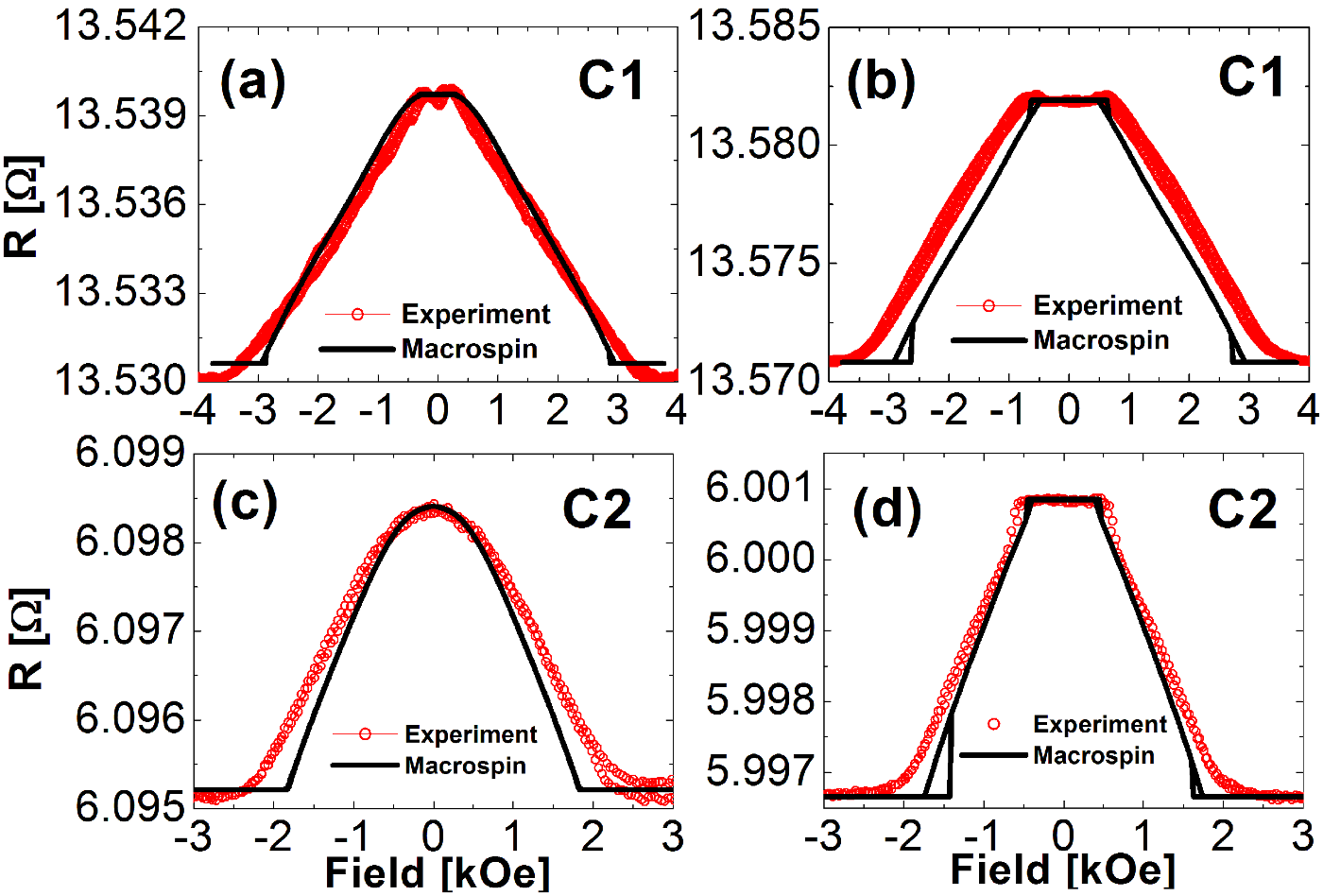}
\caption{Macrospin MR curves (black) fit to the experimental points (red) with a magnetic field applied along the hard axes of  samples (a) C1  and (c) C2  or easy axes of samples (b)\ C1  and (d) C2. 
}
\label{fig:gmrcmacro}
\end{figure}

\begin{figure*}
\includegraphics[width=12.6cm]{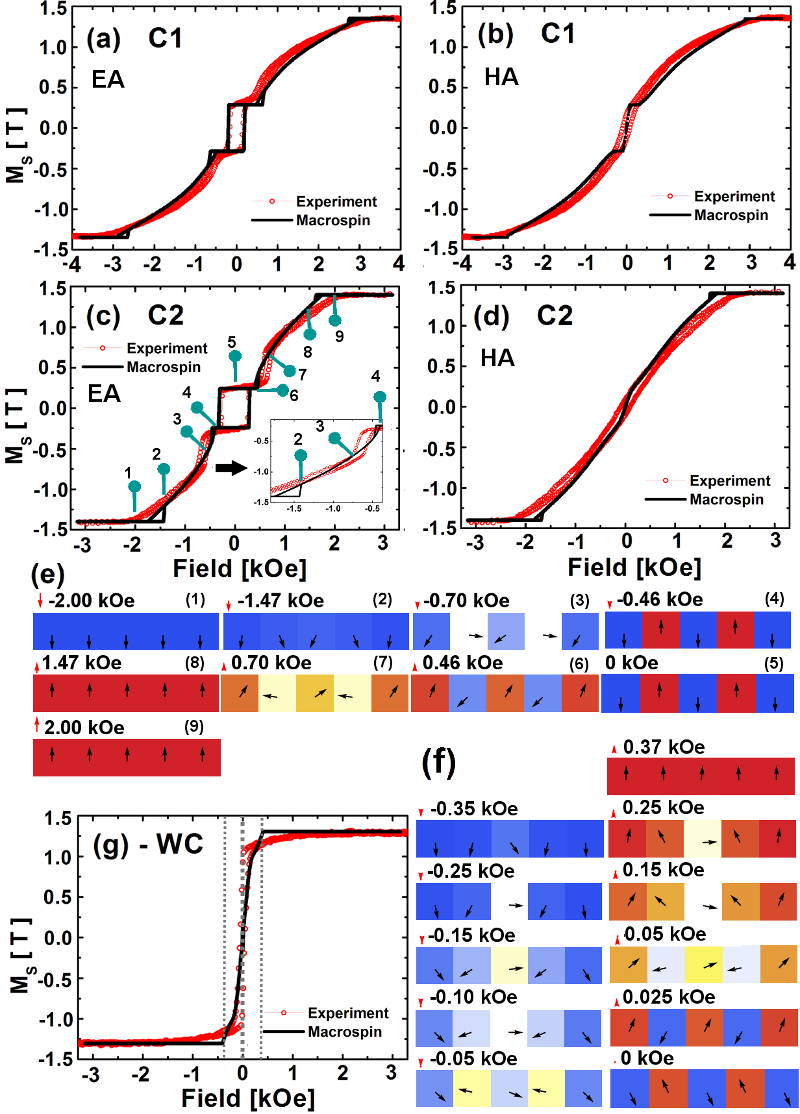}
\caption{Macrospin \(M(H)\) curves fit to the experimental dependencies measured with the magnetic field applied along the easy  and hard  axes of samples (a), (b)\ C1, (c), (d) C2, and of (g) the isotropic WC sample. The easy-axis magnetization reversal process of (e)\ C2  and (f) WC  samples as a function
of external magnetic field is illustrated by the black arrows, which indicate the magnetization direction within each Co sublayer, and the related colors: from dark blue (downward magnetization) to red (upward magnetization).  The blue thumbtack-shaped marks and the numbers on (c) the \(M(H)\)-C2 plot refer to (e)\ the respective magnetization configurations, whereas the vertical lines on \(M(H)\) plot for the (g) WC sample  indicate the magnetic fields where (f)\ the magnetization reversal occurs in the WC SAF.  
}
\label{fig:histereza}
\end{figure*}

Figures \ref{fig:gmrc} and \ref{fig:gmrcmacro} show the experimental and theoretical \(MR(H)\) dependencies measured in samples C1 and C2  and the macrospin curves, and Fig. \ref{fig:histereza} shows the \(M(H)\) relations, together with a detailed illustration of SAF magnetization reversal processes. 

In addition to the contribution from the GMR,  the magnetization orientation with   respect to the current flow  (anisotropic magnetoresistance - AMR) affects the resulting MR signal. In particular, the difference in MR measured along the easy and hard axes at saturation magnetic field, shown in Fig. \ref{fig:gmrc}, is the contribution from the AMR. The AMR increases with the thickness of the FM layers \cite{mcguire75}, so in sample C2 it is slightly greater than in sample C1.
In the field along the easy axis, the MR attains a maximum value in a certain field range [Figs. \ref{fig:gmrcmacro}(b) and \ref{fig:gmrcmacro}(d)] corresponding to the width of the central hysteresis loop in \(M(H)\) curves [Figs. \ref{fig:histereza}(a) and \ref{fig:histereza}(c)]. This is related to the stable AFM alignment of the magnetization in the component Co layers and the enhanced spin-sensitive scattering of electrons. The decrease in MR with  magnetic field corresponds to a rotation of the magnetization  toward saturation, where MR reaches its minimum value. The sloped \(R(H)\) dependencies with the apex at \(H=0\) in the field applied along the hard axis [Fig. \ref{fig:gmrcmacro}(a) and \ref{fig:gmrcmacro}(c)] correlates with the \(M(H)\) curves that makes a loopless, tilted linear-like shape  in Figs. \ref{fig:histereza}(b) and \ref{fig:histereza}(d). Such a reversible change of magnetization is expected when the magnetization of the Co component layers gradually rotates from the hard-axis direction (forced by the applied field) toward the orthogonal direction (i.e., easy axis), where the magnetization is stable in the remanent state. 

\begin{figure}[h]
\includegraphics[width=7.5cm]{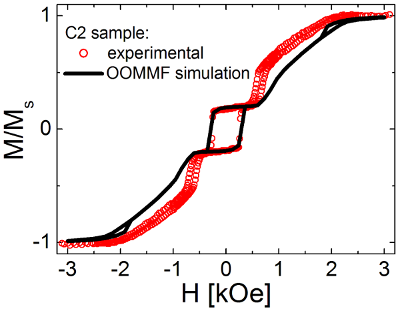}
\caption{Micromagnetically (OOMMF) simulated \(M(H)\) curve versus experimental points obtained for sample C2. See Sec.  \ref{sec:mikro} for details.}
\label{fig:mikroVSM}
\end{figure}

The experimental VSM relations measured in the C1 and C2 samples with the magnetic field applied along the easy axis exhibit a typical loop in the center of  the \(M(H)\) curve  [Figs. \ref{fig:histereza} (a) and  \ref{fig:histereza}(c)]. Upon decreasing the magnetic field, the magnetizations of the two inner Co layers rotate coherently and gradually from parallel alignment (saturated state) to  AFM ordering. The loop in the central part of the \(M(H)\) dependence is a consequence of  antiparallel magnetization configuration in the stack composed of an odd number of magnetic layers. This configuration is stable in a certain field range corresponding to the width of the hysteresis loop. A simulation of the macrospin precisely reproduces the central hysteresis loop. 
Two stable AFM orderings occur within the loop region   (3 of 5 or 2 of 5 Co-layer magnetizations aligned along the external field), so that a transition between these orderings may occur. This is seen as an abrupt switching of magnetization in individual layers [compare magnetization alignment at 0 and 0.46 kOe in Fig. \ref{fig:histereza}(e)]. Moreover, the modeling reveals another narrow subloop existing at  higher fields, close to the saturation of the SAF magnetization [see inset of Fig. \ref{fig:histereza}(c)]. In these two regions there are two energy minima that are close to each other and separated by a low potential-energy barrier. As a consequence, there are two possible magnetization configurations in these narrow ranges of magnetic field. This feature is clearly illustrated in Fig. \ref{fig:histereza}(e). In the field range corresponding to the subloops, the magnetization alignment depends on the magnetic history of the multilayer: at \(-1.4\) kOe (evolution from saturation towards AFM ordering), magnetization of the component layers is noncolinear, whereas it becomes fully colinear at +1.4 kOe (approaching saturation). Since the effect has been accounted for by the monodomain macrospin model, it can be related to the interlayer exchange coupling present in samples C1 and C2  and, to a lesser extent, to possible magnetization inhomogeneities and domain-wall pinning within the Co layers. Figure \ref{fig:mikroVSM} shows the more realistic micromagnetic simulation of the hysteresis loop measured in  sample C2 with the magnetic field applied along its easy axis (cf. Sec. \ref{sec:mikro}). Similarly to the macrospin simulations, the micromagnetics exhibits small \(M(H)\) subloops in the vicinity of the saturation field.  

The fits obtained in both macrospin and micromagnetic approaches are  satisfactory.  Therefore, the effect of the interlayer coupling  on the VSM relations  dominates over the effects related to the complex magnetization distribution within the magnetic layers.

For the WC sample,   very narrow hysteresis loops appear in both orthogonal magnetic-field directions applied in the sample plane [Fig. \ref{fig:histereza}(g)]. This is evidence that the WC sample is isotropic. In addition, the interlayer coupling strength is not sufficiently high  to produce a considerable hysteresis loop. The magnetization reversal is illustrated in Fig. \ref{fig:histereza}(f). The magnetization rotates in all magnetic layers gradually in almost the entire  range of magnetic field. The only possible switching-like event is  at \(H_{\text{ext}}\approx 0\) due to nonzero interlayer coupling. Moreover, the rotation of Co magnetization differs from that in the C2 sample. In particular, the central Co layer behaves  differently  than the others due to a smaller \(M_S\) \cite{wawro2017engineering}. The magnetization of the middle layer is softer because of its perpendicular orientation with respect to the other Co layers at relatively high magnetic field [see Fig. \ref{fig:histereza}(f) at \(H=0.25\) kOe].

The difference in the anisotropic properties of samples C1 and C2 and the isotropic properties of sample WC  can be explained in terms of strains induced in the component Co layers due to lattice mismatch at the Mo/Co interfaces. A detailed discussion of this issue is provided in  Sec. \ref{sec:struktura}.

\subsection{Ferromagnetic-resonance characterization\label{sec:dynamic}}

This section discusses the magnetic dynamics in the coupled samples versus the WC sample. By using  the VNA-FMR technique, we acquired the resonance response spectra from all  samples. Next, the relations  \(f(H_{\text{res}})\)  are extracted from the VNA spectra by fitting the relevant resonance curves. In addition, the mode intensities are modeled by using the macrospin model and micromagnetics with the same magnetic parameters as used in Sec. \ref{sec:vsm}. Again, the reliability of  the macrospin modeling is confirmed by comparing the results obtained from both approaches. The qualitative difference in \(f(H_{\text{res}})\) between coupled and WC samples is revealed and discussed. The detailed analysis of the optical and acoustic modes and their dependence on the external magnetic field in the coupled sample C2 is also provided.

\begin{figure}[h]
\includegraphics[width=8.6cm]{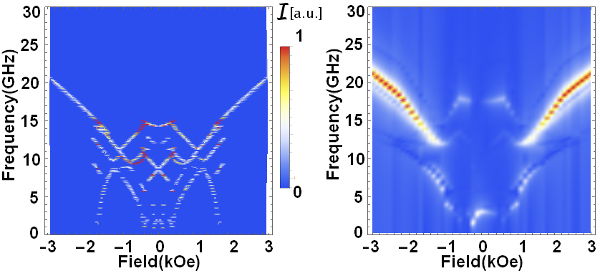}
\caption{Intensities of  resonance modes (\(\mathcal{I}\)) calculated from Eq. (\ref{eq:intense}) as a function of the resonance field \(H_{\text{res}}\) and frequency \(f\) with macrospin (left panel) and micromagnetic (right panel) approach for sample C2  with a magnetic field applied in the easy axis. The macrospin magnetic parameters are listed in Table \ref{tab:tabela}.
}
\label{fig:macromicro}
\end{figure}

Figure \ref{fig:macromicro} shows the theoretical \(f(H_{\text{res}})\) relations and mode intensities predicted for sample C2   by Eqs. (\ref{eq:czesto}) and (\ref{eq:intense}), respectively, using the parameters from Table \ref{tab:tabela}.

The C2 sample exhibits a complex spectrum of resonance modes. However, the calculated modes are influenced rather by the interlayer coupling than by the magnetization inhomogeneity within Co layers. Similar complex FMR modes have  already been observed; for example, in Fe/Cr superlattices with bi-quadratic interlayer coupling \cite{demokritovFMR} and, recently, in permalloy/Ru multilayers \cite{chaudhuri2018electrical}. The macrospin prediction was compared with that  calculated micromagnetically (cf. Sec. \ref{sec:mikro}). The comparison is shown in Fig. \ref{fig:macromicro}.

The micromagnetic and macrospin models agree well, particularly the resonance modes, which exhibit the same eigenfrequencies at high fields in both approaches. On the other hand, at low fields, the nontrivial dependence of resonance modes is accounted for by both the macrospin model and micromagnetics. The rich dispersion relation is due only to interlayer coupling, which  is justified because comparing the micromagnetic hysteresis  with the macrospin hysteresis does not reveal a significant qualitative difference [cf. Eq. (\ref{fig:mikroVSM})]. Nevertheless, the mode intensities differ in both approaches (i.e., low-field modes are more visible in macrospin than in micromagnetics). Similarly, the micromagnetic high-frequency mode at low field exhibits higher frequencies than its macrospin counterpart. 
A different number of resonance modes at low fields may be related to small amplitudes of micromagnetic oscillations. The barely visible white streaks at low resonance fields strongly suggest that such small amplitude micromagnetic oscillations are present. Therefore, we  
conclude that both approaches describe the resonance modes in the same qualitative way and the multiple resonance modes are related rather to collective magnetization dynamics of five Co exchange-coupled layers than to the magnetization inhomogeneities within them.
Also, it proves that the chosen sets of macrospin parameters are reasonable for all samples. 
The analysis of the resonance response to an AC driving magnetic field shows that the full solution of the LLG equation (2) consists of two parts: The first part corresponds to the damped free oscillations with resonance frequencies \(\omega_i\) [cf. Eq. (\ref{eq:eigensolution})]. The related terms are transient and disappear after a characteristic time \(
{\propto} \frac{1}{\omega_i}\).
The second part corresponds to the steady-state oscillations with the driving magnetic-field frequency (\(
{\propto} e^{i\omega_{\text{AC}}t}\)) \cite{pain1993}. In particular, the amplitude of this term becomes enhanced when its frequency matches the resonance frequencies \(\omega_{\text{AC}} \approx \omega_i\). Thus, the mode intensities calculated according to Eq. (\ref{eq:intense}) may not be relevant anymore, so that we calculated all resonance modes regardless of their amplitudes.   
The model predictions were compared with the experimental data. 

\begin{figure}[t]
\includegraphics[width=8.6cm]{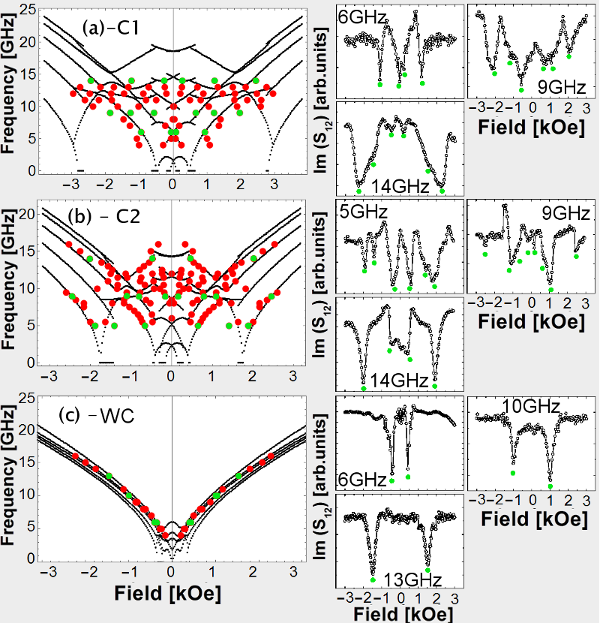}
\caption{Left: Experimental resonance fields of samples (a) C1, (b) C2, and (c) WC  determined from  VNA measurements and compared with theoretical (macrospin) dispersion relations. Right: Examples of measured VNA spectra of \(S_{12}\) parameter for  given frequencies. Green points indicate identified resonance peaks and their counterparts in the dispersion-relation plots. The VNA spectra corresponding to red points are not shown here.  
}
\label{fig:fmrall}
\end{figure}

The VNA-FMR spectra were measured by sweeping the magnetic field at constant frequency of the driving microwave magnetic field \(H_{\text{AC}}\). These measurements were repeated for several frequencies of \(H_{\text{AC}}\), ranging from 5 to 16 GHz for all samples. The experimental and theoretical results are shown in Figs. \ref{fig:fmrall}(a)--\ref{fig:fmrall}(c). 

\begin{figure}[h]
\includegraphics[width=8.6cm]{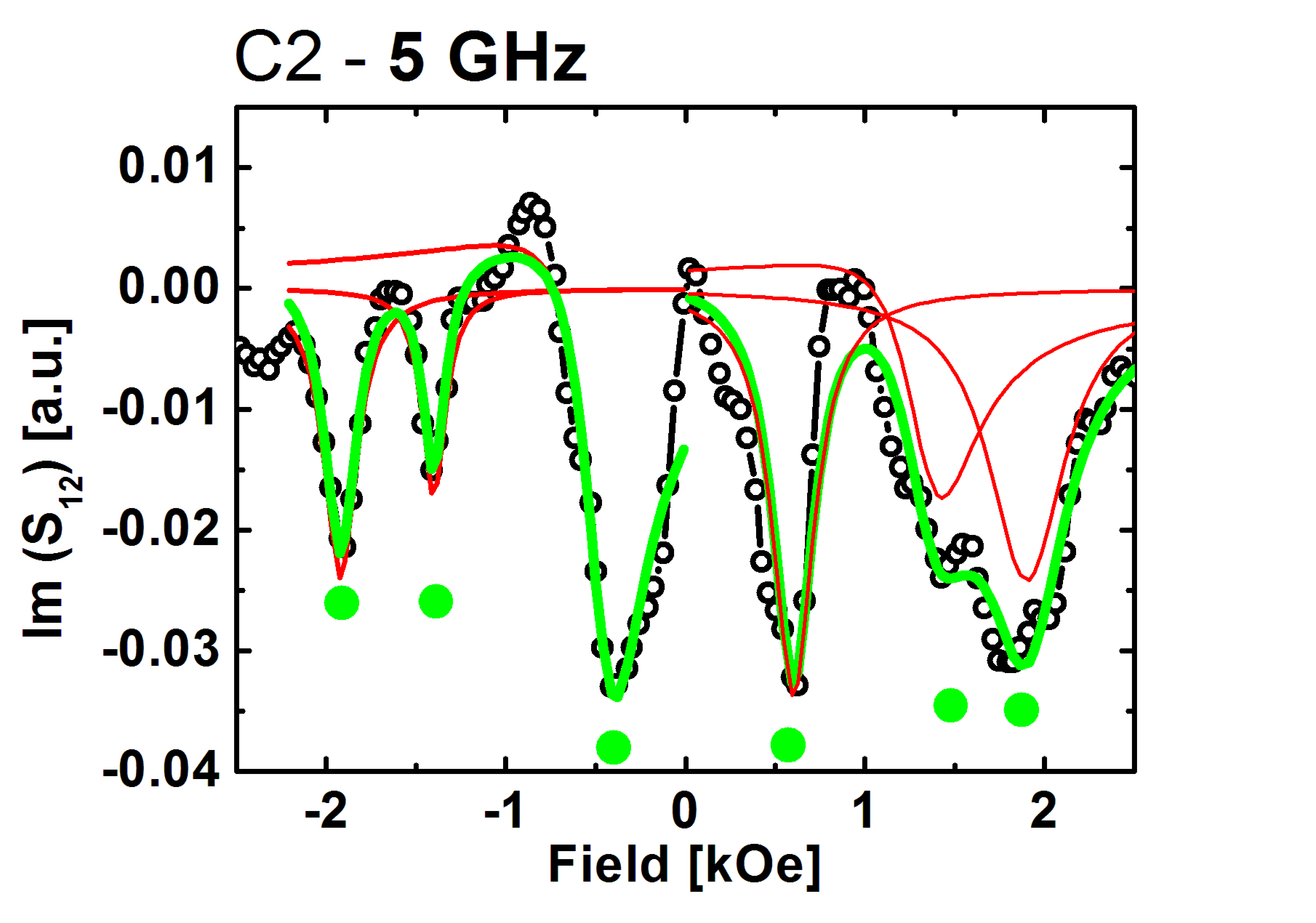}
\caption{Experimental VNA spectrum of sample C2  measured under a 5 GHz driving magnetic field. Each  separate green line is the sum of three independent Lorentzians corresponding to three distinguishable resonance peaks (red lines). The fitting procedure was done separately for positive and negative magnetic fields. 
}
\label{fig:fits}
\end{figure}

A detailed analysis reveals that, in addition to clearly visible peaks,  abrupt changes also appear in the real and imaginary part of \(S_{12}\)  for both positive and negative magnetic fields. Such  characteristic ``jumps'' can be assigned to additional resonance peaks. Also, we  identify subpeaks within the peaks with larger width. The identification of resonance peaks (resonance fields) is difficult but possible by fitting the relevant Lorentzians, including symmetric \begin{displaymath}S(H) =\frac{\Delta H^2}{(H-H_R)^2 + \Delta H^2}\end{displaymath} and antisymmetric \begin{displaymath}A(H) =\frac{\Delta H (H-H_R)}{(H-H_R)^2 + \Delta H^2}\end{displaymath}  contributions. Figure \ref{fig:fits}  shows the \(S_{12}\) spectrum of sample C2  measured at  5 GHz. 
The measured spectrum is one of the most demanding in terms of resonance-mode analysis. Because of the experimental noise, the spectrum is not fully symmetric with respect to  \(H=0\) (i.e., the amplitude, linewidth, and shape of the resonance peaks differ for positive and negative magnetic fields). For this reason Lorentzians were fit to positive and negative magnetic fields separately. Conversely, the resonance frequencies turned out to be much more symmetric, so they could be analyzed regardless of any differences in other parameters. Such an approach results in a complex relation of dispersions in the case of samples C1 and C2, and a relatively simple (Kittel-like) dependency in the case of sample WC. For the coupled samples (C1 and C2), the interlayer coupling is responsible for  splitting  the \(f(H_{\text{res}})\) branches (resonant modes) and makes the VNA spectra complex. The experimental and theoretical dispersion relations are not reducible to a simple sum of five Kittel-like dispersion relations related to five different Co layers. According to the model, the entire spectrum of resonance fields should be treated as one object, which means that every magnetic (Co) layer interacts through Mo layers with other magnetic (Co) layers, not only with the nearest magnetic neighbors. The influence of the coupling is suppressed when a Mo layer is sufficiently thick. Thus we conclude that the WC sample follows the Kittel formula, so all branches of \(f(H)\) converge to the one branch. The splitting (cf. the highest-frequency branch) visible for the WC sample is due to presence of one relatively thick Mo layer and, consequently, a very weak AFM coupling. Moreover, the middle Co layer is considerably thinner than the others. This reason is more important here because the thinness of the middle Co layer causes its \(M_S\) to be smaller as well. However, the suppressed interlayer coupling makes the VNA resonance peaks easily distinguishable. Note that, for a given frequency, most branch gaps are of the order of the VNA resonance-peak widths. Thus we cannot exclude that  more than one narrow peak exists. Such narrow peaks can be simply fit to the experimental spectra (not shown here), which is obviously justified by a model. However, here we  assume   a sufficiently wide one-resonance mode covering the five slightly split submodes. 

\begin{figure}[h]
\includegraphics[width=9.0cm]{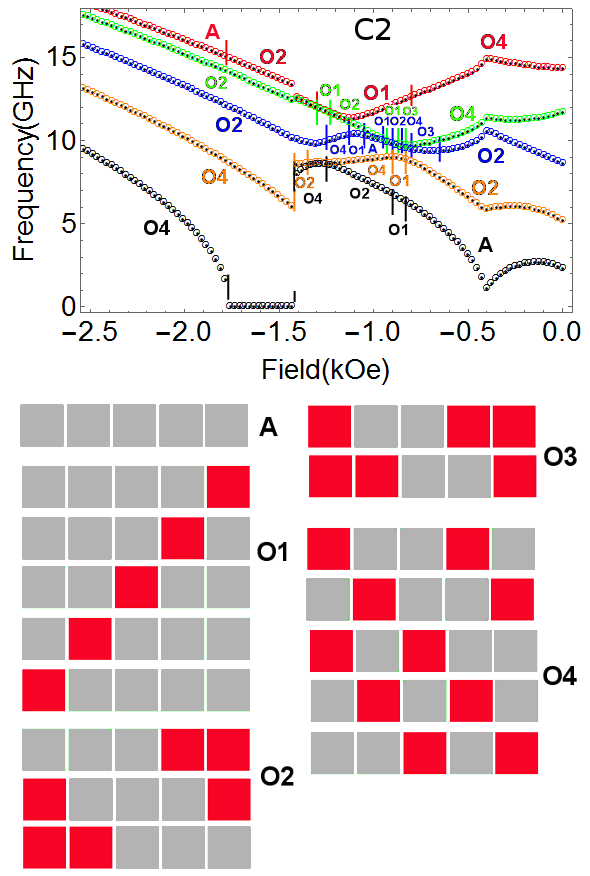}
\caption{Top: Five branches of the dispersion relation, each marked by a different color, calculated for sample C2. 
Different types of optical and acoustic modes marked  A and O1 to O4 with  colors assigned to the respective branch. A short vertical colored line indicates the transitions from one type of mode to  another within the given branch.
Bottom: Illustration of five types of oscillation modes: 1 acoustic and 4 optical modes. Five squares in a row represent the Co sublayers within the sample oscillating with the same frequency, whereas the same colors (gray or red) denote the  layers oscillating in the same phase.  Each type of modes is realized in one (in the case of A mode) or more (5,3,2 in the case of O1 and O4, O2, O3 modes, respectively) ways.
}
\label{fig:mody}
\end{figure}

A detailed analysis of oscillation-mode phase (for  sample C2) reveals a further abundance of dynamical states. Because of the number of magnetic layers, there are \(2^5\) different  in-phase and antiphase type (symmetries) of oscillation modes. By grouping  similar types of oscillations,  the number of modes reduces  to five with different symmetries denoted by letters A and O1 to O4. Here, the letters A and O stand for acoustic and optical modes, respectively; they are illustrated in Fig. \ref{fig:mody}. 
Each square corresponds to one Co layer, while its color (gray or red) indicates the phase of the magnetization oscillations, so that the same colors mean the same phase of the magnetization oscillations. 
As shown in Fig. \ref{fig:mody}, each mode has a well-established symmetry when the frequency gap between the modes is sufficiently large. In particular, it is visible in low and high magnetic fields, for which a high-frequency mode
is acoustic (optical), whereas a low-frequency mode is optical (acoustic) over a broad range of high (low) magnetic fields.  This result is consistent with  similar studies of the coupled \([\text{Fe/Cr}]\times 2 / \text{Fe}\) trilayer structure \cite{zivieri2000}. 
However, in the present case, the mode frequencies get closer to each other at intermediate magnetic fields. Moreover, for a certain range of magnetic fields, the modes frequencies overlap; that is, they synchronize their frequencies [cf. two high-frequency modes  (red and green points) the moderate modes (green and blue points) or two low-frequency modes (orange and black points)]. In the regions where the modes are close together, multiple changes are observed in their shapes (symmetries) upon  increasing the magnetic field.

\subsection{Coupling energy\label{sec:iec}}
This section parametrizes  the key relations for multilayers (i.e., the coupling energy versus Mo spacer thickness). The  relation obtained [\(J(d_{\text{Mo}})\)] for our uncompensated SAF is compared with those  reported in the literature.

\begin{figure}[h]
\includegraphics[width=8.6cm]{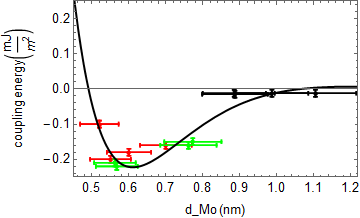}
\caption{Coupling energy as a function of  Mo layer thickness. The red, green, and black points correspond to Mo thickness determined from XRR measurement for C1, C2 and WC samples respectively. The black solid line is fit to the experimental points.}
\label{fig:coupling}
\end{figure}

Section \ref{sec:dynamic} explains that interlayer coupling is the main factor affecting the complex magnetic dynamics, and consequently samples C1 and C2  reveal qualitatively different dispersion relations than sample WC. Here, we determine how the interlayer coupling magnitude  depends on the Mo layer thickness [\(J (d_{\text{Mo}})\).  The interlayer coupling magnitude between each subsequent pair of Co sublayers (\(J_{ij}\)) can be determined as one of the fitting parameters (cf. Table \ref{tab:tabela}) and from the analysis of the XRR data (cf. Sec. \ref{sec:struktura}). The latter technique reveals the information regarding the Mo layer thickness in all samples. The plot of the semi-empirical points (i.e., \(J\) versus \(d_{\text{Mo}}\)), is shown in Fig. \ref{fig:coupling}. The thickness \(d_{\text{Mo}}\) has  a maximum relative error of 10\%, while the absolute error of \(J_{ij}\) is estimated to be \(\pm 0.01\text{ mJ/m}^2\). Such a value for the absolute error ensures that the experimental dependencies of VSM, VNA, and MR are satisfactorily reproduced by the theoretical model. 

The theoretical curve was fit to the semi-empirical points by using the  RKKY-like interaction \cite{kurant2019}:
\begin{equation}
J(d_{\text{Mo}}) = \frac{A}{d_{\text{Mo}}^2}\sin \left( 2 \pi \frac{d_{\text{Mo}}}{\Lambda} + \psi \right) e^{\frac{d_{\text{Mo}}}{t_c}},
\label{eq:rkky}
\end{equation} 
where \(A\) and \(\psi\) are the coupling amplitude and phase, respectively, and \(\Lambda\) and \(t_c\) are the period of the coupling oscillations and attenuation length, respectively. The fitting to the semi-empirical points was done with the following parameter values: \(A=1.4\text{ J}\), \(\Lambda=1.15\text{ nm}\), \(\psi=0.45\), \(t_c = 0.26\text{ nm}\). 
\begin{table}[h!]
\centering
\small
\caption{Comparison of  interlayer coupling parameters for different Co/Mo structures: five Co/Mo bilayers (present work), Co/Mo wedge (Ref. [\onlinecite{kurant2019}]) and 16 Co/Mo bilayers (Ref. [\onlinecite{parkinCoMo}]). \(d_{\text{Mo,min}}^{\text{AFM}}\): Mo thickness at which the coupling dependence exhibits a first AFM minimum, \(J (d_{\text{Mo,min}}^{\text{AFM}})\): magnitude of coupling at first AFM minimum, \(\Lambda\): period of coupling oscillations,  \(\Delta d_{\text{Mo,1}}^{\text{AFM}}\) - range of Mo thickness where the coupling is AFM for the first AFM minimum.
 \label{tab:tabela3}}
\begin{tabular}{ccccc}
\hline\hline
\text{Structure } &  \(d_{\text{Mo,min}}^{\text{AFM}}  \) & \( J(d_{\text{Mo,min}}^{\text{AFM}})  \)  &  \( \Lambda \)  &  \(\Delta d_{\text{Mo,1}}^{\text{AFM}} \) \\
 &   [nm] &  \([\frac{\text{J}}{\text{m}^2}]\)  &   [nm]  &   [nm]\\ 
\hline
  \text{Present work}  & 0.61  & \(-0.22\)  & 1.15  & 0.57   \\
  \text{Ref. [\onlinecite{kurant2019}]}  & 0.7  & \(-0.33\)  & 1.40 & 0.71   \\
  \text{Ref. [\onlinecite{parkinCoMo}]}  & 0.52  & \(-0.12\)  & 1.10  & 0.30   \\
  \hline\hline
\end{tabular}
\end{table}
The result is shown as a solid line in Fig. \ref{fig:coupling}. A deep minimum around \(d_{\text{Mo}} \approx 0.6\) nm reflects the maximum AFM strength of the interlayer coupling. The obtained parameters are reliable and close to those reported for Co/Mo/Co trilayers \cite{kurant2019} and for the system consisting of 16 Co/Mo bilayers \cite{parkinCoMo}. To compare the results obtained herein with those from Refs. [\onlinecite{kurant2019,parkinCoMo}], we converted our parameters  into the same quantities as used in these references. Their comparison is given in Table \ref{tab:tabela3}. Our parameters are located in the middle between the wedge structure and multilayered Co/Mo structure. This supports our argument for the reliability of the parametrization of the coupling.

\section{Summary and conclusions\label{sec:sumary}}
Epitaxially grown Co/Mo multilayers exhibit diverse magnetic properties. The interlayer AFM or FM coupling  can be tuned by varying the thickness of the Mo spacer. The superlattices analyzed herein reveal substantial AFM interlayer coupling strength within certain ranges of Mo spacer thickness, making them applicable as uncompensated SAFs (cf. Sec. \ref{sec:iec}). Moreover, the symmetry and lattice parameter mismatches at the interfaces generate anisotropic strain in the Co layers and, consequently, due to magnetoelastic effects,  in-plane twofold magnetic anisotropy is induced. This result allows the mutually orthogonal easy and hard axes of magnetization to be easily determined (cf. Sec. \ref{sec:struktura}). Therefore, in such structures, the magnetization orientation in the sample plane is well defined, which is important for applications. 
These complex magnetic properties are clearly explained by detailed analyses of the structural properties obtained from complementary techniques (XRD, XRR, and TEM), which are presented in Sec. \ref{sec:struktura}.
Moreover, the magnetic configuration of the multilayer substantially affects the magnetization-reversal process (cf. Sec. \ref{sec:vsm}). 
%and dynamic behavior making FMR dispersion spectra complex or simple for AFM or FM alignment, respectively (cf. Sec\ref{sec:dynamic}). 
Results of numerical modeling based on macrospin and micromagnetic approaches reproduce well the experimental results and  allow us to determine a reliable set of magnetic parameters (cf. Sec. \ref{sec:vsm}) describing the Co/Mo multilayers. In particular, we  show that  considering the real thickness of component layers is of significant importance for the analysis of magnetodynamic and magnetostatic properties of Co/Mo superlattices. Such a detailed approach allows us to obtain the interlayer-coupling dependence on Mo-spacer thickness (cf. Sec. \ref{sec:iec}). The coupling is of the RKKY type and is similar to that for Co/Mo sandwiches. Despite a relatively strong interlayer coupling, the MR in multilayers, which is similar to that of a bilayer system, is very weak (cf. Sec. \ref{sec:vsm}). 
The reliable parametrization of the interlayer coupling and of the magnetic-layer properties  reveals qualitative differences in dispersion relations in coupled versus WC SAFs (cf. Sec. \ref{sec:dynamic}). Finally, the acoustic and optical modes and their evolution with resonance magnetic field were indentified in a strongly coupled SAF (cf. Sec. \ref{sec:dynamic}). 

\section*{ACKNOWLEDGMENTS\label{sec:acknow}}
This work was financed by the National Science Centre in Poland under the project: SPINORBITRONICS 2016/23/B/ST3/01430 (acknowledgments from J.K., P.O.,  S.Z., M.C., and T.S.) and 2014/13/B/ST5/01834 (acknowledgments from A.W., J.K.,  A.P., K.M., P.D., and K.D.) and co-financed by the EU European Regional Development Fund (REINTEGRATION 2017 OPIE 14-20) (acknowledgments from A.W. and A.P.). Numerical calculations were supported in part by the PL-GRID infrastructure. All authors thank S. \L{}azarski and M. Matczak for their help with VSM and MR measurements.

\bibliography{bibliography}

\begin{thebibliography}{52}
\expandafter\ifx\csname natexlab\endcsname\relax\def\natexlab#1{#1}\fi
\expandafter\ifx\csname bibnamefont\endcsname\relax
  \def\bibnamefont#1{#1}\fi
\expandafter\ifx\csname bibfnamefont\endcsname\relax
  \def\bibfnamefont#1{#1}\fi
\expandafter\ifx\csname citenamefont\endcsname\relax
  \def\citenamefont#1{#1}\fi
\expandafter\ifx\csname url\endcsname\relax
  \def\url#1{\texttt{#1}}\fi
\expandafter\ifx\csname urlprefix\endcsname\relax\def\urlprefix{URL }\fi
\providecommand{\bibinfo}[2]{#2}
\providecommand{\eprint}[2][]{\url{#2}}

\bibitem[{\citenamefont{Gr{\"u}nberg et~al.}(1986)\citenamefont{Gr{\"u}nberg,
  Schreiber, Pang, Brodsky, and Sowers}}]{grunberg1986layered}
\bibinfo{author}{\bibfnamefont{P.}~\bibnamefont{Gr{\"u}nberg}},
  \bibinfo{author}{\bibfnamefont{R.}~\bibnamefont{Schreiber}},
  \bibinfo{author}{\bibfnamefont{Y.}~\bibnamefont{Pang}},
  \bibinfo{author}{\bibfnamefont{M.~B.} \bibnamefont{Brodsky}},
  \bibnamefont{and} \bibinfo{author}{\bibfnamefont{H.}~\bibnamefont{Sowers}},
  \bibinfo{journal}{Phys. Rev. Lett.} \textbf{\bibinfo{volume}{57}},
  \bibinfo{pages}{2442} (\bibinfo{year}{1986}),
  \urlprefix\url{https://link.aps.org/doi/10.1103/PhysRevLett.57.2442}.

\bibitem[{\citenamefont{Majkrzak et~al.}(1986)\citenamefont{Majkrzak, Cable,
  Kwo, Hong, McWhan, Yafet, Waszczak, and Vettier}}]{majkrzak1986cf}
\bibinfo{author}{\bibfnamefont{C.~F.} \bibnamefont{Majkrzak}},
  \bibinfo{author}{\bibfnamefont{J.~W.} \bibnamefont{Cable}},
  \bibinfo{author}{\bibfnamefont{J.}~\bibnamefont{Kwo}},
  \bibinfo{author}{\bibfnamefont{M.}~\bibnamefont{Hong}},
  \bibinfo{author}{\bibfnamefont{D.~B.} \bibnamefont{McWhan}},
  \bibinfo{author}{\bibfnamefont{Y.}~\bibnamefont{Yafet}},
  \bibinfo{author}{\bibfnamefont{J.~V.} \bibnamefont{Waszczak}},
  \bibnamefont{and} \bibinfo{author}{\bibfnamefont{C.}~\bibnamefont{Vettier}},
  \bibinfo{journal}{Phys. Rev. Lett.} \textbf{\bibinfo{volume}{56}},
  \bibinfo{pages}{2700} (\bibinfo{year}{1986}),
  \urlprefix\url{https://link.aps.org/doi/10.1103/PhysRevLett.56.2700}.

\bibitem[{\citenamefont{Salamon et~al.}(1986)\citenamefont{Salamon, Sinha,
  Rhyne, Cunningham, Erwin, Borchers, and Flynn}}]{salamon1986long}
\bibinfo{author}{\bibfnamefont{M.~B.} \bibnamefont{Salamon}},
  \bibinfo{author}{\bibfnamefont{S.}~\bibnamefont{Sinha}},
  \bibinfo{author}{\bibfnamefont{J.~J.} \bibnamefont{Rhyne}},
  \bibinfo{author}{\bibfnamefont{J.~E.} \bibnamefont{Cunningham}},
  \bibinfo{author}{\bibfnamefont{R.~W.} \bibnamefont{Erwin}},
  \bibinfo{author}{\bibfnamefont{J.}~\bibnamefont{Borchers}}, \bibnamefont{and}
  \bibinfo{author}{\bibfnamefont{C.~P.} \bibnamefont{Flynn}},
  \bibinfo{journal}{Phys. Rev. Lett.} \textbf{\bibinfo{volume}{56}},
  \bibinfo{pages}{259} (\bibinfo{year}{1986}),
  \urlprefix\url{https://link.aps.org/doi/10.1103/PhysRevLett.56.259}.

\bibitem[{\citenamefont{Carbone and Alvarado}(1987)}]{carbone1987antiparallel}
\bibinfo{author}{\bibfnamefont{C.}~\bibnamefont{Carbone}} \bibnamefont{and}
  \bibinfo{author}{\bibfnamefont{S.~F.} \bibnamefont{Alvarado}},
  \bibinfo{journal}{Phys. Rev. B} \textbf{\bibinfo{volume}{36}},
  \bibinfo{pages}{2433} (\bibinfo{year}{1987}),
  \urlprefix\url{https://link.aps.org/doi/10.1103/PhysRevB.36.2433}.

\bibitem[{\citenamefont{Grolier et~al.}(1993)\citenamefont{Grolier, Renard,
  Bartenlian, Beauvillain, Chappert, Dupas, Ferr{\'e}, Galtier, Kolb, Mulloy
  et~al.}}]{grolier1993unambiguous}
\bibinfo{author}{\bibfnamefont{V.}~\bibnamefont{Grolier}},
  \bibinfo{author}{\bibfnamefont{D.}~\bibnamefont{Renard}},
  \bibinfo{author}{\bibfnamefont{B.}~\bibnamefont{Bartenlian}},
  \bibinfo{author}{\bibfnamefont{P.}~\bibnamefont{Beauvillain}},
  \bibinfo{author}{\bibfnamefont{C.}~\bibnamefont{Chappert}},
  \bibinfo{author}{\bibfnamefont{C.}~\bibnamefont{Dupas}},
  \bibinfo{author}{\bibfnamefont{J.}~\bibnamefont{Ferr{\'e}}},
  \bibinfo{author}{\bibfnamefont{M.}~\bibnamefont{Galtier}},
  \bibinfo{author}{\bibfnamefont{E.}~\bibnamefont{Kolb}},
  \bibinfo{author}{\bibfnamefont{M.}~\bibnamefont{Mulloy}},
  \bibnamefont{et~al.}, \bibinfo{journal}{Phys. Rev. Lett.}
  \textbf{\bibinfo{volume}{71}}, \bibinfo{pages}{3023} (\bibinfo{year}{1993}),
  \urlprefix\url{https://link.aps.org/doi/10.1103/PhysRevLett.71.3023}.

\bibitem[{\citenamefont{Bruno and Chappert}(1991)}]{bruno1991oscillatory}
\bibinfo{author}{\bibfnamefont{P.}~\bibnamefont{Bruno}} \bibnamefont{and}
  \bibinfo{author}{\bibfnamefont{C.}~\bibnamefont{Chappert}},
  \bibinfo{journal}{Phys. Rev. Lett.} \textbf{\bibinfo{volume}{67}},
  \bibinfo{pages}{1602} (\bibinfo{year}{1991}),
  \urlprefix\url{https://link.aps.org/doi/10.1103/PhysRevLett.67.1602}.

\bibitem[{\citenamefont{Bruno}(1995)}]{bruno1995theory}
\bibinfo{author}{\bibfnamefont{P.}~\bibnamefont{Bruno}},
  \bibinfo{journal}{Phys. Rev. B} \textbf{\bibinfo{volume}{52}},
  \bibinfo{pages}{411} (\bibinfo{year}{1995}),
  \urlprefix\url{https://link.aps.org/doi/10.1103/PhysRevB.52.411}.

\bibitem[{\citenamefont{Baibich et~al.}(1988)\citenamefont{Baibich, Broto,
  Fert, Van~Dau, Petroff, Etienne, Creuzet, Friederich, and
  Chazelas}}]{baibich1988giant}
\bibinfo{author}{\bibfnamefont{M.~N.} \bibnamefont{Baibich}},
  \bibinfo{author}{\bibfnamefont{J.~M.} \bibnamefont{Broto}},
  \bibinfo{author}{\bibfnamefont{A.}~\bibnamefont{Fert}},
  \bibinfo{author}{\bibfnamefont{F.~N.} \bibnamefont{Van~Dau}},
  \bibinfo{author}{\bibfnamefont{F.}~\bibnamefont{Petroff}},
  \bibinfo{author}{\bibfnamefont{P.}~\bibnamefont{Etienne}},
  \bibinfo{author}{\bibfnamefont{G.}~\bibnamefont{Creuzet}},
  \bibinfo{author}{\bibfnamefont{A.}~\bibnamefont{Friederich}},
  \bibnamefont{and} \bibinfo{author}{\bibfnamefont{J.}~\bibnamefont{Chazelas}},
  \bibinfo{journal}{Phys. Rev. Lett.} \textbf{\bibinfo{volume}{61}},
  \bibinfo{pages}{2472} (\bibinfo{year}{1988}),
  \urlprefix\url{https://link.aps.org/doi/10.1103/PhysRevLett.61.2472}.

\bibitem[{\citenamefont{Camley and Barna{\'s}}(1989)}]{camley1989theory}
\bibinfo{author}{\bibfnamefont{R.~E.} \bibnamefont{Camley}} \bibnamefont{and}
  \bibinfo{author}{\bibfnamefont{J.}~\bibnamefont{Barna{\'s}}},
  \bibinfo{journal}{Phys. Rev. Lett.} \textbf{\bibinfo{volume}{63}},
  \bibinfo{pages}{664} (\bibinfo{year}{1989}),
  \urlprefix\url{https://link.aps.org/doi/10.1103/PhysRevLett.63.664}.

\bibitem[{\citenamefont{Gomez-Perez et~al.}(2018)\citenamefont{Gomez-Perez,
  V{\'e}lez, McKenzie-Sell, Amado, Herrero-Mart{\'{\i}n}, L{\'o}pez-L{\'o}pez,
  Blanco-Canosa, Hueso, Chuvilin, Robinson et~al.}}]{gomez2018synthetic}
\bibinfo{author}{\bibfnamefont{J.~M.} \bibnamefont{Gomez-Perez}},
  \bibinfo{author}{\bibfnamefont{S.}~\bibnamefont{V{\'e}lez}},
  \bibinfo{author}{\bibfnamefont{L.}~\bibnamefont{McKenzie-Sell}},
  \bibinfo{author}{\bibfnamefont{M.}~\bibnamefont{Amado}},
  \bibinfo{author}{\bibfnamefont{J.}~\bibnamefont{Herrero-Mart{\'{\i}n}}},
  \bibinfo{author}{\bibfnamefont{J.}~\bibnamefont{L{\'o}pez-L{\'o}pez}},
  \bibinfo{author}{\bibfnamefont{S.}~\bibnamefont{Blanco-Canosa}},
  \bibinfo{author}{\bibfnamefont{L.~E.} \bibnamefont{Hueso}},
  \bibinfo{author}{\bibfnamefont{A.}~\bibnamefont{Chuvilin}},
  \bibinfo{author}{\bibfnamefont{J.~W.~A.} \bibnamefont{Robinson}},
  \bibnamefont{et~al.}, \bibinfo{journal}{Phys. Rev. Applied}
  \textbf{\bibinfo{volume}{10}}, \bibinfo{pages}{044046}
  (\bibinfo{year}{2018}),
  \urlprefix\url{https://link.aps.org/doi/10.1103/PhysRevApplied.10.044046}.

\bibitem[{\citenamefont{Liu et~al.}(2018)\citenamefont{Liu, Wu, Couet, Mertens,
  Rao, Kim, Garello, Crotti, Van~Elshocht, De~Boeck et~al.}}]{liu2018synthetic}
\bibinfo{author}{\bibfnamefont{E.}~\bibnamefont{Liu}},
  \bibinfo{author}{\bibfnamefont{Y.-C.} \bibnamefont{Wu}},
  \bibinfo{author}{\bibfnamefont{S.}~\bibnamefont{Couet}},
  \bibinfo{author}{\bibfnamefont{S.}~\bibnamefont{Mertens}},
  \bibinfo{author}{\bibfnamefont{S.}~\bibnamefont{Rao}},
  \bibinfo{author}{\bibfnamefont{W.}~\bibnamefont{Kim}},
  \bibinfo{author}{\bibfnamefont{K.}~\bibnamefont{Garello}},
  \bibinfo{author}{\bibfnamefont{D.}~\bibnamefont{Crotti}},
  \bibinfo{author}{\bibfnamefont{S.}~\bibnamefont{Van~Elshocht}},
  \bibinfo{author}{\bibfnamefont{J.}~\bibnamefont{De~Boeck}},
  \bibnamefont{et~al.}, \bibinfo{journal}{Phys. Rev. Applied}
  \textbf{\bibinfo{volume}{10}}, \bibinfo{pages}{054054}
  (\bibinfo{year}{2018}),
  \urlprefix\url{https://link.aps.org/doi/10.1103/PhysRevApplied.10.054054}.

\bibitem[{\citenamefont{Chatterjee et~al.}(2018)\citenamefont{Chatterjee,
  Auffret, Sousa, Coelho, Prejbeanu, and Dieny}}]{chatterjee2018novel}
\bibinfo{author}{\bibfnamefont{J.}~\bibnamefont{Chatterjee}},
  \bibinfo{author}{\bibfnamefont{S.}~\bibnamefont{Auffret}},
  \bibinfo{author}{\bibfnamefont{R.}~\bibnamefont{Sousa}},
  \bibinfo{author}{\bibfnamefont{P.}~\bibnamefont{Coelho}},
  \bibinfo{author}{\bibfnamefont{I.-L.} \bibnamefont{Prejbeanu}},
  \bibnamefont{and} \bibinfo{author}{\bibfnamefont{B.}~\bibnamefont{Dieny}},
  \bibinfo{journal}{Sci. Rep.} \textbf{\bibinfo{volume}{8}},
  \bibinfo{pages}{11724} (\bibinfo{year}{2018}),
  \urlprefix\url{https://doi.org/10.1038/s41598-018-29913-6}.

\bibitem[{\citenamefont{Chaudhuri et~al.}(2018)\citenamefont{Chaudhuri, Xiong,
  Mahendiran, and Adeyeye}}]{chaudhuri2018electrical}
\bibinfo{author}{\bibfnamefont{U.}~\bibnamefont{Chaudhuri}},
  \bibinfo{author}{\bibfnamefont{L.}~\bibnamefont{Xiong}},
  \bibinfo{author}{\bibfnamefont{R.}~\bibnamefont{Mahendiran}},
  \bibnamefont{and} \bibinfo{author}{\bibfnamefont{A.~O.}
  \bibnamefont{Adeyeye}}, \bibinfo{journal}{Appl. Phys. Lett.}
  \textbf{\bibinfo{volume}{113}}, \bibinfo{pages}{262406}
  (\bibinfo{year}{2018}), \urlprefix\url{http://doi.org/10.1063/1.5054358}.

\bibitem[{\citenamefont{Zhang et~al.}(2018)\citenamefont{Zhang, Liao, Shi,
  Zhang, Wu, Wang, Pan, and Song}}]{zhangSOT}
\bibinfo{author}{\bibfnamefont{P.~X.} \bibnamefont{Zhang}},
  \bibinfo{author}{\bibfnamefont{L.~Y.} \bibnamefont{Liao}},
  \bibinfo{author}{\bibfnamefont{G.~Y.} \bibnamefont{Shi}},
  \bibinfo{author}{\bibfnamefont{R.~Q.} \bibnamefont{Zhang}},
  \bibinfo{author}{\bibfnamefont{H.~Q.} \bibnamefont{Wu}},
  \bibinfo{author}{\bibfnamefont{Y.~Y.} \bibnamefont{Wang}},
  \bibinfo{author}{\bibfnamefont{F.}~\bibnamefont{Pan}}, \bibnamefont{and}
  \bibinfo{author}{\bibfnamefont{C.}~\bibnamefont{Song}},
  \bibinfo{journal}{Phys. Rev. B} \textbf{\bibinfo{volume}{97}},
  \bibinfo{pages}{214403} (\bibinfo{year}{2018}),
  \urlprefix\url{https://link.aps.org/doi/10.1103/PhysRevB.97.214403}.

\bibitem[{\citenamefont{Wu et~al.}(2019)\citenamefont{Wu, Razavi, Shao, Li,
  Wong, Liu, Yin, and Wang}}]{wuSOT}
\bibinfo{author}{\bibfnamefont{H.}~\bibnamefont{Wu}},
  \bibinfo{author}{\bibfnamefont{S.~A.} \bibnamefont{Razavi}},
  \bibinfo{author}{\bibfnamefont{Q.}~\bibnamefont{Shao}},
  \bibinfo{author}{\bibfnamefont{X.}~\bibnamefont{Li}},
  \bibinfo{author}{\bibfnamefont{K.~L.} \bibnamefont{Wong}},
  \bibinfo{author}{\bibfnamefont{Y.}~\bibnamefont{Liu}},
  \bibinfo{author}{\bibfnamefont{G.}~\bibnamefont{Yin}}, \bibnamefont{and}
  \bibinfo{author}{\bibfnamefont{K.~L.} \bibnamefont{Wang}},
  \bibinfo{journal}{Phys. Rev. B} \textbf{\bibinfo{volume}{99}},
  \bibinfo{pages}{184403} (\bibinfo{year}{2019}),
  \urlprefix\url{https://link.aps.org/doi/10.1103/PhysRevB.99.184403}.

\bibitem[{\citenamefont{Farle}(1998)}]{farle1998ferromagnetic}
\bibinfo{author}{\bibfnamefont{M.}~\bibnamefont{Farle}}, \bibinfo{journal}{Rep.
  Prog. Phys.} \textbf{\bibinfo{volume}{61}}, \bibinfo{pages}{755}
  (\bibinfo{year}{1998}),
  \urlprefix\url{https://doi.org/10.1088%2F0034-4885%2F61%2F7%2F001}.

\bibitem[{\citenamefont{Lindner and
  Baberschke}(2003)}]{lindner2003ferromagnetic}
\bibinfo{author}{\bibfnamefont{J.}~\bibnamefont{Lindner}} \bibnamefont{and}
  \bibinfo{author}{\bibfnamefont{K.}~\bibnamefont{Baberschke}},
  \bibinfo{journal}{J. Phys. Condens. Mat.} \textbf{\bibinfo{volume}{15}},
  \bibinfo{pages}{S465} (\bibinfo{year}{2003}),
  \urlprefix\url{https://doi.org/10.1088%2F0953-8984%2F15%2F5%2F303}.

\bibitem[{\citenamefont{Khodadadi et~al.}(2017)\citenamefont{Khodadadi,
  Mohammadi, Jones, Srivastava, Mewes, Mewes, and
  Kaiser}}]{khodadadi2017interlayer}
\bibinfo{author}{\bibfnamefont{B.}~\bibnamefont{Khodadadi}},
  \bibinfo{author}{\bibfnamefont{J.~B.} \bibnamefont{Mohammadi}},
  \bibinfo{author}{\bibfnamefont{J.~M.} \bibnamefont{Jones}},
  \bibinfo{author}{\bibfnamefont{A.}~\bibnamefont{Srivastava}},
  \bibinfo{author}{\bibfnamefont{C.}~\bibnamefont{Mewes}},
  \bibinfo{author}{\bibfnamefont{T.}~\bibnamefont{Mewes}}, \bibnamefont{and}
  \bibinfo{author}{\bibfnamefont{C.}~\bibnamefont{Kaiser}},
  \bibinfo{journal}{Phys. Rev. Applied} \textbf{\bibinfo{volume}{8}},
  \bibinfo{pages}{014024} (\bibinfo{year}{2017}),
  \urlprefix\url{https://link.aps.org/doi/10.1103/PhysRevApplied.8.014024}.

\bibitem[{\citenamefont{Rementer et~al.}(2017)\citenamefont{Rementer, Fitzell,
  Xu, Nordeen, Carman, Wang, and Chang}}]{rementer2017tuning}
\bibinfo{author}{\bibfnamefont{C.~R.} \bibnamefont{Rementer}},
  \bibinfo{author}{\bibfnamefont{K.}~\bibnamefont{Fitzell}},
  \bibinfo{author}{\bibfnamefont{Q.}~\bibnamefont{Xu}},
  \bibinfo{author}{\bibfnamefont{P.}~\bibnamefont{Nordeen}},
  \bibinfo{author}{\bibfnamefont{G.~P.} \bibnamefont{Carman}},
  \bibinfo{author}{\bibfnamefont{Y.~E.} \bibnamefont{Wang}}, \bibnamefont{and}
  \bibinfo{author}{\bibfnamefont{J.~P.} \bibnamefont{Chang}},
  \bibinfo{journal}{Appl. Phys. Lett.} \textbf{\bibinfo{volume}{110}},
  \bibinfo{pages}{242403} (\bibinfo{year}{2017}),
  \urlprefix\url{https://doi.org/10.1063/1.4984298}.

\bibitem[{\citenamefont{Bezerra et~al.}(1999)\citenamefont{Bezerra,
  de~Ara\'ujo, Chesman, and Albuquerque}}]{bezerra1999self}
\bibinfo{author}{\bibfnamefont{C.~G.} \bibnamefont{Bezerra}},
  \bibinfo{author}{\bibfnamefont{J.~M.} \bibnamefont{de~Ara\'ujo}},
  \bibinfo{author}{\bibfnamefont{C.}~\bibnamefont{Chesman}}, \bibnamefont{and}
  \bibinfo{author}{\bibfnamefont{E.~L.} \bibnamefont{Albuquerque}},
  \bibinfo{journal}{Phys. Rev. B} \textbf{\bibinfo{volume}{60}},
  \bibinfo{pages}{9264} (\bibinfo{year}{1999}),
  \urlprefix\url{https://link.aps.org/doi/10.1103/PhysRevB.60.9264}.

\bibitem[{\citenamefont{Sousa et~al.}(2019)\citenamefont{Sousa,
  Quispe-Marcatoma, Pelegrini, and Baggio-Saitovitch}}]{sousa2019ferromagnetic}
\bibinfo{author}{\bibfnamefont{M.}~\bibnamefont{Sousa}},
  \bibinfo{author}{\bibfnamefont{J.}~\bibnamefont{Quispe-Marcatoma}},
  \bibinfo{author}{\bibfnamefont{F.}~\bibnamefont{Pelegrini}},
  \bibnamefont{and}
  \bibinfo{author}{\bibfnamefont{E.}~\bibnamefont{Baggio-Saitovitch}},
  \bibinfo{journal}{J. Magn. Magn. Mater.} \textbf{\bibinfo{volume}{474}},
  \bibinfo{pages}{250} (\bibinfo{year}{2019}),
  \urlprefix\url{http://doi.org/10.1016/j.jmmm.2018.10.127}.

\bibitem[{\citenamefont{Li et~al.}(2018)\citenamefont{Li, Miao, Cao, Li, Xu,
  Wen, Dai, Yan, and L{\"u}}}]{li2018stress}
\bibinfo{author}{\bibfnamefont{S.}~\bibnamefont{Li}},
  \bibinfo{author}{\bibfnamefont{G.-X.} \bibnamefont{Miao}},
  \bibinfo{author}{\bibfnamefont{D.}~\bibnamefont{Cao}},
  \bibinfo{author}{\bibfnamefont{Q.}~\bibnamefont{Li}},
  \bibinfo{author}{\bibfnamefont{J.}~\bibnamefont{Xu}},
  \bibinfo{author}{\bibfnamefont{Z.}~\bibnamefont{Wen}},
  \bibinfo{author}{\bibfnamefont{Y.}~\bibnamefont{Dai}},
  \bibinfo{author}{\bibfnamefont{S.}~\bibnamefont{Yan}}, \bibnamefont{and}
  \bibinfo{author}{\bibfnamefont{Y.}~\bibnamefont{L{\"u}}},
  \bibinfo{journal}{ACS Appl. Mater. Interfaces} \textbf{\bibinfo{volume}{10}},
  \bibinfo{pages}{8853} (\bibinfo{year}{2018}),
  \urlprefix\url{https://doi.org/10.1021/acsami.7b19684}.

\bibitem[{\citenamefont{Hrabec et~al.}(2014)\citenamefont{Hrabec, Porter,
  Wells, Benitez, Burnell, McVitie, McGrouther, Moore, and
  Marrows}}]{hrabec2014}
\bibinfo{author}{\bibfnamefont{A.}~\bibnamefont{Hrabec}},
  \bibinfo{author}{\bibfnamefont{N.~A.} \bibnamefont{Porter}},
  \bibinfo{author}{\bibfnamefont{A.}~\bibnamefont{Wells}},
  \bibinfo{author}{\bibfnamefont{M.~J.} \bibnamefont{Benitez}},
  \bibinfo{author}{\bibfnamefont{G.}~\bibnamefont{Burnell}},
  \bibinfo{author}{\bibfnamefont{S.}~\bibnamefont{McVitie}},
  \bibinfo{author}{\bibfnamefont{D.}~\bibnamefont{McGrouther}},
  \bibinfo{author}{\bibfnamefont{T.~A.} \bibnamefont{Moore}}, \bibnamefont{and}
  \bibinfo{author}{\bibfnamefont{C.~H.} \bibnamefont{Marrows}},
  \bibinfo{journal}{Phys. Rev. B} \textbf{\bibinfo{volume}{90}},
  \bibinfo{pages}{020402} (\bibinfo{year}{2014}),
  \urlprefix\url{https://link.aps.org/doi/10.1103/PhysRevB.90.020402}.

\bibitem[{\citenamefont{Fert et~al.}(2017)\citenamefont{Fert, Reyren, and
  Cros}}]{fert2017magnetic}
\bibinfo{author}{\bibfnamefont{A.}~\bibnamefont{Fert}},
  \bibinfo{author}{\bibfnamefont{N.}~\bibnamefont{Reyren}}, \bibnamefont{and}
  \bibinfo{author}{\bibfnamefont{V.}~\bibnamefont{Cros}},
  \bibinfo{journal}{Nature Reviews Materials} \textbf{\bibinfo{volume}{2}},
  \bibinfo{pages}{17031} (\bibinfo{year}{2017}),
  \urlprefix\url{https://doi.org/10.1038/natrevmats.2016.44}.

\bibitem[{\citenamefont{Zhao et~al.}(2003)\citenamefont{Zhao, Yang, Zeng, and
  Pan}}]{zhao2003irradiation}
\bibinfo{author}{\bibfnamefont{B.}~\bibnamefont{Zhao}},
  \bibinfo{author}{\bibfnamefont{G.}~\bibnamefont{Yang}},
  \bibinfo{author}{\bibfnamefont{F.}~\bibnamefont{Zeng}}, \bibnamefont{and}
  \bibinfo{author}{\bibfnamefont{F.}~\bibnamefont{Pan}}, \bibinfo{journal}{Acta
  Mater.} \textbf{\bibinfo{volume}{51}}, \bibinfo{pages}{5093}
  (\bibinfo{year}{2003}),
  \urlprefix\url{https://doi.org/10.1016/S1359-6454(03)00359-8}.

\bibitem[{\citenamefont{Yang et~al.}(2004)\citenamefont{Yang, Gu, Gao, and
  Pan}}]{yang2004evolution}
\bibinfo{author}{\bibfnamefont{G.}~\bibnamefont{Yang}},
  \bibinfo{author}{\bibfnamefont{Y.}~\bibnamefont{Gu}},
  \bibinfo{author}{\bibfnamefont{Y.}~\bibnamefont{Gao}}, \bibnamefont{and}
  \bibinfo{author}{\bibfnamefont{F.}~\bibnamefont{Pan}}, \bibinfo{journal}{Thin
  Solid Films} \textbf{\bibinfo{volume}{457}}, \bibinfo{pages}{354}
  (\bibinfo{year}{2004}),
  \urlprefix\url{https://doi.org/10.1016/j.tsf.2003.11.290}.

\bibitem[{\citenamefont{Yang and Pan}(2002)}]{yang2002structural}
\bibinfo{author}{\bibfnamefont{G.}~\bibnamefont{Yang}} \bibnamefont{and}
  \bibinfo{author}{\bibfnamefont{F.}~\bibnamefont{Pan}}, \bibinfo{journal}{J.
  Magn. Magn. Mater.} \textbf{\bibinfo{volume}{250}}, \bibinfo{pages}{249}
  (\bibinfo{year}{2002}),
  \urlprefix\url{https://doi.org/10.1016/S0304-8853(02)00402-X}.

\bibitem[{\citenamefont{Wang et~al.}(1991)\citenamefont{Wang, Cui, Li, and
  Fan}}]{wang1991co}
\bibinfo{author}{\bibfnamefont{Y.}~\bibnamefont{Wang}},
  \bibinfo{author}{\bibfnamefont{F.}~\bibnamefont{Cui}},
  \bibinfo{author}{\bibfnamefont{W.}~\bibnamefont{Li}}, \bibnamefont{and}
  \bibinfo{author}{\bibfnamefont{Y.}~\bibnamefont{Fan}}, \bibinfo{journal}{J.
  Magn. Magn. Mater.} \textbf{\bibinfo{volume}{102}}, \bibinfo{pages}{121}
  (\bibinfo{year}{1991}),
  \urlprefix\url{https://doi.org/10.1016/0304-8853(91)90276-G}.

\bibitem[{\citenamefont{Houserov{\'a} et~al.}(2005)\citenamefont{Houserov{\'a},
  V{\v{r}}e{\v{s}}t'{\'a}l, and {\v{S}}ob}}]{houserova2005phase}
\bibinfo{author}{\bibfnamefont{J.}~\bibnamefont{Houserov{\'a}}},
  \bibinfo{author}{\bibfnamefont{J.}~\bibnamefont{V{\v{r}}e{\v{s}}t'{\'a}l}},
  \bibnamefont{and}
  \bibinfo{author}{\bibfnamefont{M.}~\bibnamefont{{\v{S}}ob}},
  \bibinfo{journal}{Calphad} \textbf{\bibinfo{volume}{29}},
  \bibinfo{pages}{133} (\bibinfo{year}{2005}),
  \urlprefix\url{https://doi.org/10.1016/j.calphad.2005.06.002}.

\bibitem[{\citenamefont{Guo et~al.}(2003)\citenamefont{Guo, Kong, Liu, and
  Liu}}]{guo2003ab}
\bibinfo{author}{\bibfnamefont{H.}~\bibnamefont{Guo}},
  \bibinfo{author}{\bibfnamefont{L.}~\bibnamefont{Kong}},
  \bibinfo{author}{\bibfnamefont{J.}~\bibnamefont{Liu}}, \bibnamefont{and}
  \bibinfo{author}{\bibfnamefont{B.}~\bibnamefont{Liu}},
  \bibinfo{journal}{Solid State Commun.} \textbf{\bibinfo{volume}{125}},
  \bibinfo{pages}{435} (\bibinfo{year}{2003}),
  \urlprefix\url{https://doi.org/10.1016/S0038-1098(02)00824-4}.

\bibitem[{\citenamefont{Parkin}(1991)}]{parkinCoMo}
\bibinfo{author}{\bibfnamefont{S.~S.~P.} \bibnamefont{Parkin}},
  \bibinfo{journal}{Phys. Rev. Lett.} \textbf{\bibinfo{volume}{67}},
  \bibinfo{pages}{3598} (\bibinfo{year}{1991}),
  \urlprefix\url{https://link.aps.org/doi/10.1103/PhysRevLett.67.3598}.

\bibitem[{\citenamefont{Shalyguina et~al.}(2006)\citenamefont{Shalyguina,
  Perepelova, Kozlovskii, and Tamanis}}]{shalyguina2006magneto}
\bibinfo{author}{\bibfnamefont{E.}~\bibnamefont{Shalyguina}},
  \bibinfo{author}{\bibfnamefont{E.}~\bibnamefont{Perepelova}},
  \bibinfo{author}{\bibfnamefont{L.}~\bibnamefont{Kozlovskii}},
  \bibnamefont{and} \bibinfo{author}{\bibfnamefont{E.}~\bibnamefont{Tamanis}},
  \bibinfo{journal}{J. Magn. Magn. Mater.} \textbf{\bibinfo{volume}{300}},
  \bibinfo{pages}{e363} (\bibinfo{year}{2006}),
  \urlprefix\url{http://www.sciencedirect.com/science/article/pii/S0304885305009704}.

\bibitem[{\citenamefont{Koelling}(1994)}]{koelling1994magnetic}
\bibinfo{author}{\bibfnamefont{D.~D.} \bibnamefont{Koelling}},
  \bibinfo{journal}{Phys. Rev. B} \textbf{\bibinfo{volume}{50}},
  \bibinfo{pages}{273} (\bibinfo{year}{1994}),
  \urlprefix\url{https://link.aps.org/doi/10.1103/PhysRevB.50.273}.

\bibitem[{\citenamefont{Kurant et~al.}(2019)\citenamefont{Kurant, Tekielak,
  Sveklo, Wawro, and Maziewski}}]{kurant2019}
\bibinfo{author}{\bibfnamefont{Z.}~\bibnamefont{Kurant}},
  \bibinfo{author}{\bibfnamefont{M.}~\bibnamefont{Tekielak}},
  \bibinfo{author}{\bibfnamefont{I.}~\bibnamefont{Sveklo}},
  \bibinfo{author}{\bibfnamefont{A.}~\bibnamefont{Wawro}}, \bibnamefont{and}
  \bibinfo{author}{\bibfnamefont{A.}~\bibnamefont{Maziewski}},
  \bibinfo{journal}{J. Magn. Magn. Mater.} \textbf{\bibinfo{volume}{475}},
  \bibinfo{pages}{683 } (\bibinfo{year}{2019}), ISSN \bibinfo{issn}{0304-8853},
  \urlprefix\url{http://www.sciencedirect.com/science/article/pii/S0304885318335984}.

\bibitem[{\citenamefont{Wawro et~al.}(2017{\natexlab{a}})\citenamefont{Wawro,
  Kurant, Tekielak, Nawrocki, Mili{\'n}ska, Pietruczik, W{\'o}jcik, Mazalski,
  Kanak, Ollefs et~al.}}]{wawro2017engineering}
\bibinfo{author}{\bibfnamefont{A.}~\bibnamefont{Wawro}},
  \bibinfo{author}{\bibfnamefont{Z.}~\bibnamefont{Kurant}},
  \bibinfo{author}{\bibfnamefont{M.}~\bibnamefont{Tekielak}},
  \bibinfo{author}{\bibfnamefont{P.}~\bibnamefont{Nawrocki}},
  \bibinfo{author}{\bibfnamefont{E.}~\bibnamefont{Mili{\'n}ska}},
  \bibinfo{author}{\bibfnamefont{A.}~\bibnamefont{Pietruczik}},
  \bibinfo{author}{\bibfnamefont{M.}~\bibnamefont{W{\'o}jcik}},
  \bibinfo{author}{\bibfnamefont{P.}~\bibnamefont{Mazalski}},
  \bibinfo{author}{\bibfnamefont{J.}~\bibnamefont{Kanak}},
  \bibinfo{author}{\bibfnamefont{K.}~\bibnamefont{Ollefs}},
  \bibnamefont{et~al.}, \bibinfo{journal}{J. Phys. D Appl. Phys.}
  \textbf{\bibinfo{volume}{50}}, \bibinfo{pages}{215004}
  (\bibinfo{year}{2017}{\natexlab{a}}),
  \urlprefix\url{https://doi.org/10.1088/1361-6463/aa6a94}.

\bibitem[{\citenamefont{Wawro et~al.}(2017{\natexlab{b}})\citenamefont{Wawro,
  Kurant, Tekielak, Jakubowski, Pietruczik, B{\"o}ttger, and
  Maziewski}}]{wawro2017modifications}
\bibinfo{author}{\bibfnamefont{A.}~\bibnamefont{Wawro}},
  \bibinfo{author}{\bibfnamefont{Z.}~\bibnamefont{Kurant}},
  \bibinfo{author}{\bibfnamefont{M.}~\bibnamefont{Tekielak}},
  \bibinfo{author}{\bibfnamefont{M.}~\bibnamefont{Jakubowski}},
  \bibinfo{author}{\bibfnamefont{A.}~\bibnamefont{Pietruczik}},
  \bibinfo{author}{\bibfnamefont{R.}~\bibnamefont{B{\"o}ttger}},
  \bibnamefont{and}
  \bibinfo{author}{\bibfnamefont{A.}~\bibnamefont{Maziewski}},
  \bibinfo{journal}{Appl. Phys. Lett.} \textbf{\bibinfo{volume}{110}},
  \bibinfo{pages}{252405} (\bibinfo{year}{2017}{\natexlab{b}}),
  \urlprefix\url{https://doi.org/10.1063/1.4987142}.

\bibitem[{\citenamefont{Wawro et~al.}(2018{\natexlab{a}})\citenamefont{Wawro,
  Kurant, Jakubowski, Tekielak, Pietruczik, B\"ottger, and
  Maziewski}}]{wawro2018magnetic}
\bibinfo{author}{\bibfnamefont{A.}~\bibnamefont{Wawro}},
  \bibinfo{author}{\bibfnamefont{Z.}~\bibnamefont{Kurant}},
  \bibinfo{author}{\bibfnamefont{M.}~\bibnamefont{Jakubowski}},
  \bibinfo{author}{\bibfnamefont{M.}~\bibnamefont{Tekielak}},
  \bibinfo{author}{\bibfnamefont{A.}~\bibnamefont{Pietruczik}},
  \bibinfo{author}{\bibfnamefont{R.}~\bibnamefont{B\"ottger}},
  \bibnamefont{and}
  \bibinfo{author}{\bibfnamefont{A.}~\bibnamefont{Maziewski}},
  \bibinfo{journal}{Phys. Rev. Applied} \textbf{\bibinfo{volume}{9}},
  \bibinfo{pages}{014029} (\bibinfo{year}{2018}{\natexlab{a}}),
  \urlprefix\url{https://link.aps.org/doi/10.1103/PhysRevApplied.9.014029}.

\bibitem[{\citenamefont{Wawro et~al.}(2018{\natexlab{b}})\citenamefont{Wawro,
  Mili{\'n}ska, Kurant, Pietruczik, Kanak, Ollefs, Wilhelm, Rogalev, and
  Maziewski}}]{wawro2018momag}
\bibinfo{author}{\bibfnamefont{A.}~\bibnamefont{Wawro}},
  \bibinfo{author}{\bibfnamefont{E.}~\bibnamefont{Mili{\'n}ska}},
  \bibinfo{author}{\bibfnamefont{Z.}~\bibnamefont{Kurant}},
  \bibinfo{author}{\bibfnamefont{A.}~\bibnamefont{Pietruczik}},
  \bibinfo{author}{\bibfnamefont{J.}~\bibnamefont{Kanak}},
  \bibinfo{author}{\bibfnamefont{K.}~\bibnamefont{Ollefs}},
  \bibinfo{author}{\bibfnamefont{F.}~\bibnamefont{Wilhelm}},
  \bibinfo{author}{\bibfnamefont{A.}~\bibnamefont{Rogalev}}, \bibnamefont{and}
  \bibinfo{author}{\bibfnamefont{A.}~\bibnamefont{Maziewski}},
  \bibinfo{journal}{J. Synchrotron Rad.} \textbf{\bibinfo{volume}{25}},
  \bibinfo{pages}{1400} (\bibinfo{year}{2018}{\natexlab{b}}),
  \urlprefix\url{https://doi.org/10.1107/S1600577518008500}.

\bibitem[{\citenamefont{Sveklo et~al.}(2019)\citenamefont{Sveklo, Kurant,
  Tekielak, Pietruczik, Dybko, Wawro, and A.}}]{ukrsvekloi2019}
\bibinfo{author}{\bibfnamefont{I.}~\bibnamefont{Sveklo}},
  \bibinfo{author}{\bibfnamefont{Z.}~\bibnamefont{Kurant}},
  \bibinfo{author}{\bibfnamefont{M.}~\bibnamefont{Tekielak}},
  \bibinfo{author}{\bibfnamefont{A.}~\bibnamefont{Pietruczik}},
  \bibinfo{author}{\bibfnamefont{K.}~\bibnamefont{Dybko}},
  \bibinfo{author}{\bibfnamefont{A.}~\bibnamefont{Wawro}}, \bibnamefont{and}
  \bibinfo{author}{\bibfnamefont{M.}~\bibnamefont{A.}}, \bibinfo{journal}{J.
  Magn. Magn. Mater.} \textbf{\bibinfo{volume}{489}}, \bibinfo{pages}{165417}
  (\bibinfo{year}{2019}),
  \urlprefix\url{http://www.sciencedirect.com/science/article/pii/S0304885319304238}.

\bibitem[{\citenamefont{Donahue and Porter}(1999)}]{donahue1999oommf}
\bibinfo{author}{\bibfnamefont{M.~J.} \bibnamefont{Donahue}} \bibnamefont{and}
  \bibinfo{author}{\bibfnamefont{D.~G.} \bibnamefont{Porter}},
  \emph{\bibinfo{title}{OOMMF User's guide}} (\bibinfo{publisher}{US Department
  of Commerce, Technology Administration, National Institute of Standards and
  Technology}, \bibinfo{year}{1999}).

\bibitem[{\citenamefont{Kanak et~al.}(2013)\citenamefont{Kanak, Wiśniowski,
  Stobiecki, Zaleski, Powroźnik, Cardoso, and Freitas}}]{kanak2013}
\bibinfo{author}{\bibfnamefont{J.}~\bibnamefont{Kanak}},
  \bibinfo{author}{\bibfnamefont{P.}~\bibnamefont{Wiśniowski}},
  \bibinfo{author}{\bibfnamefont{T.}~\bibnamefont{Stobiecki}},
  \bibinfo{author}{\bibfnamefont{A.}~\bibnamefont{Zaleski}},
  \bibinfo{author}{\bibfnamefont{W.}~\bibnamefont{Powroźnik}},
  \bibinfo{author}{\bibfnamefont{S.}~\bibnamefont{Cardoso}}, \bibnamefont{and}
  \bibinfo{author}{\bibfnamefont{P.~P.} \bibnamefont{Freitas}},
  \bibinfo{journal}{J. Appl. Phys.} \textbf{\bibinfo{volume}{113}},
  \bibinfo{pages}{023915} (\bibinfo{year}{2013}),
  \eprint{https://doi.org/10.1063/1.4775706},
  \urlprefix\url{https://doi.org/10.1063/1.4775706}.

\bibitem[{\citenamefont{Prokop et~al.}(2004)\citenamefont{Prokop, Valdaitsev,
  Kukunin, Pratzer, Sch\"onhense, and Elmers}}]{prokop2004}
\bibinfo{author}{\bibfnamefont{J.}~\bibnamefont{Prokop}},
  \bibinfo{author}{\bibfnamefont{D.~A.} \bibnamefont{Valdaitsev}},
  \bibinfo{author}{\bibfnamefont{A.}~\bibnamefont{Kukunin}},
  \bibinfo{author}{\bibfnamefont{M.}~\bibnamefont{Pratzer}},
  \bibinfo{author}{\bibfnamefont{G.}~\bibnamefont{Sch\"onhense}},
  \bibnamefont{and} \bibinfo{author}{\bibfnamefont{H.~J.}
  \bibnamefont{Elmers}}, \bibinfo{journal}{Phys. Rev. B}
  \textbf{\bibinfo{volume}{70}}, \bibinfo{pages}{184423}
  (\bibinfo{year}{2004}),
  \urlprefix\url{https://link.aps.org/doi/10.1103/PhysRevB.70.184423}.

\bibitem[{\citenamefont{Smit and Beljers}(1955)}]{smit1955}
\bibinfo{author}{\bibfnamefont{J.}~\bibnamefont{Smit}} \bibnamefont{and}
  \bibinfo{author}{\bibfnamefont{H.}~\bibnamefont{Beljers}},
  \bibinfo{journal}{Philips Res. Rep.} \textbf{\bibinfo{volume}{10}},
  \bibinfo{pages}{1113} (\bibinfo{year}{1955}).

\bibitem[{\citenamefont{Barati et~al.}(2014)\citenamefont{Barati, Cinal,
  Edwards, and Umerski}}]{barati}
\bibinfo{author}{\bibfnamefont{E.}~\bibnamefont{Barati}},
  \bibinfo{author}{\bibfnamefont{M.}~\bibnamefont{Cinal}},
  \bibinfo{author}{\bibfnamefont{D.~M.} \bibnamefont{Edwards}},
  \bibnamefont{and} \bibinfo{author}{\bibfnamefont{A.}~\bibnamefont{Umerski}},
  \bibinfo{journal}{Phys. Rev. B} \textbf{\bibinfo{volume}{90}},
  \bibinfo{pages}{014420} (\bibinfo{year}{2014}),
  \urlprefix\url{https://link.aps.org/doi/10.1103/PhysRevB.90.014420}.

\bibitem[{\citenamefont{Ogrodnik et~al.}(2018)\citenamefont{Ogrodnik,
  Vetr{\`{o}}, Frankowski, Ch{\k{e}}ci{\'{n}}ski, Stobiecki, Barna{\'{s}}, and
  Ansermet}}]{Ogrodnik_2018}
\bibinfo{author}{\bibfnamefont{P.}~\bibnamefont{Ogrodnik}},
  \bibinfo{author}{\bibfnamefont{F.~A.} \bibnamefont{Vetr{\`{o}}}},
  \bibinfo{author}{\bibfnamefont{M.}~\bibnamefont{Frankowski}},
  \bibinfo{author}{\bibfnamefont{J.}~\bibnamefont{Ch{\k{e}}ci{\'{n}}ski}},
  \bibinfo{author}{\bibfnamefont{T.}~\bibnamefont{Stobiecki}},
  \bibinfo{author}{\bibfnamefont{J.}~\bibnamefont{Barna{\'{s}}}},
  \bibnamefont{and} \bibinfo{author}{\bibfnamefont{J.-P.}
  \bibnamefont{Ansermet}}, \bibinfo{journal}{J. Phys. D Appl. Phys.}
  \textbf{\bibinfo{volume}{52}}, \bibinfo{pages}{065002}
  (\bibinfo{year}{2018}),
  \urlprefix\url{https://doi.org/10.1088%2F1361-6463%2Faaf093}.

\bibitem[{\citenamefont{Frankowski et~al.}(2015)\citenamefont{Frankowski,
  Ch{\k{e}}ci{\'{n}}ski, and Czapkiewicz}}]{frankowskicomputer}
\bibinfo{author}{\bibfnamefont{M.}~\bibnamefont{Frankowski}},
  \bibinfo{author}{\bibfnamefont{J.}~\bibnamefont{Ch{\k{e}}ci{\'{n}}ski}},
  \bibnamefont{and}
  \bibinfo{author}{\bibfnamefont{M.}~\bibnamefont{Czapkiewicz}},
  \bibinfo{journal}{Computer Physics Communications}
  \textbf{\bibinfo{volume}{189}}, \bibinfo{pages}{207 } (\bibinfo{year}{2015}),
  ISSN \bibinfo{issn}{0010-4655},
  \urlprefix\url{http://www.sciencedirect.com/science/article/pii/S0010465514004159}.

\bibitem[{\citenamefont{Bloemen}(1996)}]{bloemen1996interlayer}
\bibinfo{author}{\bibfnamefont{P.}~\bibnamefont{Bloemen}},
  \bibinfo{journal}{Acta Phys. Pol. A} \textbf{\bibinfo{volume}{89}},
  \bibinfo{pages}{277} (\bibinfo{year}{1996}).

\bibitem[{\citenamefont{Shukh et~al.}(1994)\citenamefont{Shukh, Shin, and
  Hoffmann}}]{shukh1994dependence}
\bibinfo{author}{\bibfnamefont{A.}~\bibnamefont{Shukh}},
  \bibinfo{author}{\bibfnamefont{D.}~\bibnamefont{Shin}}, \bibnamefont{and}
  \bibinfo{author}{\bibfnamefont{H.}~\bibnamefont{Hoffmann}},
  \bibinfo{journal}{J. Appl. Phys.} \textbf{\bibinfo{volume}{76}},
  \bibinfo{pages}{6507} (\bibinfo{year}{1994}),
  \urlprefix\url{https://doi.org/10.1063/1.358244}.

\bibitem[{\citenamefont{McGuire and Potter}(1975)}]{mcguire75}
\bibinfo{author}{\bibfnamefont{T.}~\bibnamefont{McGuire}} \bibnamefont{and}
  \bibinfo{author}{\bibfnamefont{R.}~\bibnamefont{Potter}},
  \bibinfo{journal}{IEEE Trans. Mag.} \textbf{\bibinfo{volume}{11}},
  \bibinfo{pages}{1018} (\bibinfo{year}{1975}),
  \urlprefix\url{https://doi.org/10.1109/TMAG.1975.1058782}.

\bibitem[{\citenamefont{Drovosekov et~al.}(2001)\citenamefont{Drovosekov,
  Kholin, Kreines, Zhotikova, and Demokritov}}]{demokritovFMR}
\bibinfo{author}{\bibfnamefont{A.}~\bibnamefont{Drovosekov}},
  \bibinfo{author}{\bibfnamefont{D.}~\bibnamefont{Kholin}},
  \bibinfo{author}{\bibfnamefont{N.}~\bibnamefont{Kreines}},
  \bibinfo{author}{\bibfnamefont{O.}~\bibnamefont{Zhotikova}},
  \bibnamefont{and}
  \bibinfo{author}{\bibfnamefont{S.}~\bibnamefont{Demokritov}},
  \bibinfo{journal}{Journal of Magnetism and Magnetic Materials}
  \textbf{\bibinfo{volume}{226-230}}, \bibinfo{pages}{1779 }
  (\bibinfo{year}{2001}), ISSN \bibinfo{issn}{0304-8853},
  \bibinfo{note}{proceedings of the International Conference on Magnetism (ICM
  2000)},
  \urlprefix\url{http://www.sciencedirect.com/science/article/pii/S0304885300010131}.

\bibitem[{\citenamefont{Pain and Beyer}(1993)}]{pain1993}
\bibinfo{author}{\bibfnamefont{H.~J.} \bibnamefont{Pain}} \bibnamefont{and}
  \bibinfo{author}{\bibfnamefont{R.~T.} \bibnamefont{Beyer}},
  \emph{\bibinfo{title}{The physics of vibrations and waves}}
  (\bibinfo{publisher}{ASA}, \bibinfo{year}{1993}).

\bibitem[{\citenamefont{Zivieri et~al.}(2000)\citenamefont{Zivieri, Giovannini,
  and Nizzoli}}]{zivieri2000}
\bibinfo{author}{\bibfnamefont{R.}~\bibnamefont{Zivieri}},
  \bibinfo{author}{\bibfnamefont{L.}~\bibnamefont{Giovannini}},
  \bibnamefont{and} \bibinfo{author}{\bibfnamefont{F.}~\bibnamefont{Nizzoli}},
  \bibinfo{journal}{Phys. Rev. B} \textbf{\bibinfo{volume}{62}},
  \bibinfo{pages}{14950} (\bibinfo{year}{2000}),
  \urlprefix\url{https://link.aps.org/doi/10.1103/PhysRevB.62.14950}.

\end{thebibliography}
\end{document}